\documentclass{ieeeaccess}
\usepackage{cite}
\usepackage{amsmath,amssymb,amsfonts}
\usepackage{algorithmic}
\usepackage{graphicx}
\usepackage{array}
\usepackage{caption}
\usepackage{subfigure}

\usepackage{multirow}

\usepackage{url}

\newcommand{\widthone}{0.19}

\newcommand{\widthtwo}{0.30}

\usepackage{textcomp}
\def\BibTeX{{\rm B\kern-.05em{\sc i\kern-.025em b}\kern-.08em
    T\kern-.1667em\lower.7ex\hbox{E}\kern-.125emX}}
\begin{document}

\title{sZoom: A Framework for Automatic Zoom into High Resolution Surveillance Videos}
\author{\uppercase{Mukesh Saini}\authorrefmark{1}, 
\uppercase{Benjamin Guthier\authorrefmark{2}, Hao Kuang\authorrefmark{3}, Dwarikanath Mahapatra \authorrefmark{4} and Abdulmotaleb El Saddik}.\authorrefmark{3},
}
\address[1]{ Indian Institute of Technology Ropar (e-mail: mukesh@iitrpr.ac.in)}
\address[2]{University of Mannheim (e-mail: bguthier@pi4.informatik.uni-mannheim.de)}
\address[3]{University of Ottawa (e-mail: hkuan041@uottawa.ca, elsaddik@uottawa.ca)}
\address[4]{IBM Research - Australia, Melbourne (e-mail: dwarim@au1.ibm.com)}



\corresp{Corresponding author: Mukesh Saini (e-mail: mukesh@iitrpr.ac.in).}

\begin{abstract}

Current cameras are capable of recording high resolution video. While viewing on a mobile device, a user can manually zoom into this high resolution video to get more detailed view of objects and activities. However, manual zooming is not suitable for surveillance and monitoring. It is tiring to continuously keep zooming into various regions of the video. Also, while viewing one region, the operator may miss activities in other regions.   In this paper, we propose sZoom, a framework to automatically zoom into a high resolution surveillance video. The proposed framework selectively zooms into the sensitive regions of the video to present details of the scene, while still preserving the overall context required for situation assessment. A multi-variate Gaussian penalty is introduced to ensure full coverage of the scene. The method achieves near real-time performance through a number of timing optimizations. An extensive user study shows that, while watching a full HD video on a mobile device, the system enhances the security operator's efficiency in understanding the details of the scene by 99\% on the average compared to a scaled version of the original high resolution video. The produced video achieved 46\% higher ratings for usefulness in a surveillance task.
\end{abstract}


\titlepgskip=-15pt

\maketitle

\section{Introduction} 
The cameras employed in current surveillance systems are capable of recording high resolution video covering a large area, such as a parking lot. While the high resolution video has detailed information of the captured area, the information is only accessible with very large display device, which is expensive and not feasible in many cases. It is difficult for surveillance operators to grasp the details when looking at the full view of the video on a traditional display device such as mobile or desktop. Alternatively, there are systems that allow the users to manually zoom into the video to get more details \cite{shafiei2012jiku}. Manual zooming, however, is not suitable for surveillance application becuase of the following two regions:
\begin{itemize}
\item  Manual zooming requires operators to continuously keep zooming into different areas to obtain full coverage of the scene, which is tiring.
\item While viewing a specific region, the other regions become opaque to the operator and there is a risk of missing essential information. 
\end{itemize}




Therefore, with emerging sensing technologies, there is a need for system that can automatically zoom into the surveillance videos for security operators. However, there are a set of challenges that need to be addressed in order to realize an effective zoom functionality for surveillance videos. The first challenge is to identify important or sensitive regions from the noisy semantic observations. The second challenge is to ensure sufficient coverage to provide context for situation assessment. And the third challenge is to consistently and naturally zoom onto non-static sensitive targets.    
  
\begin{figure}
\centering
\includegraphics[width=1\linewidth]{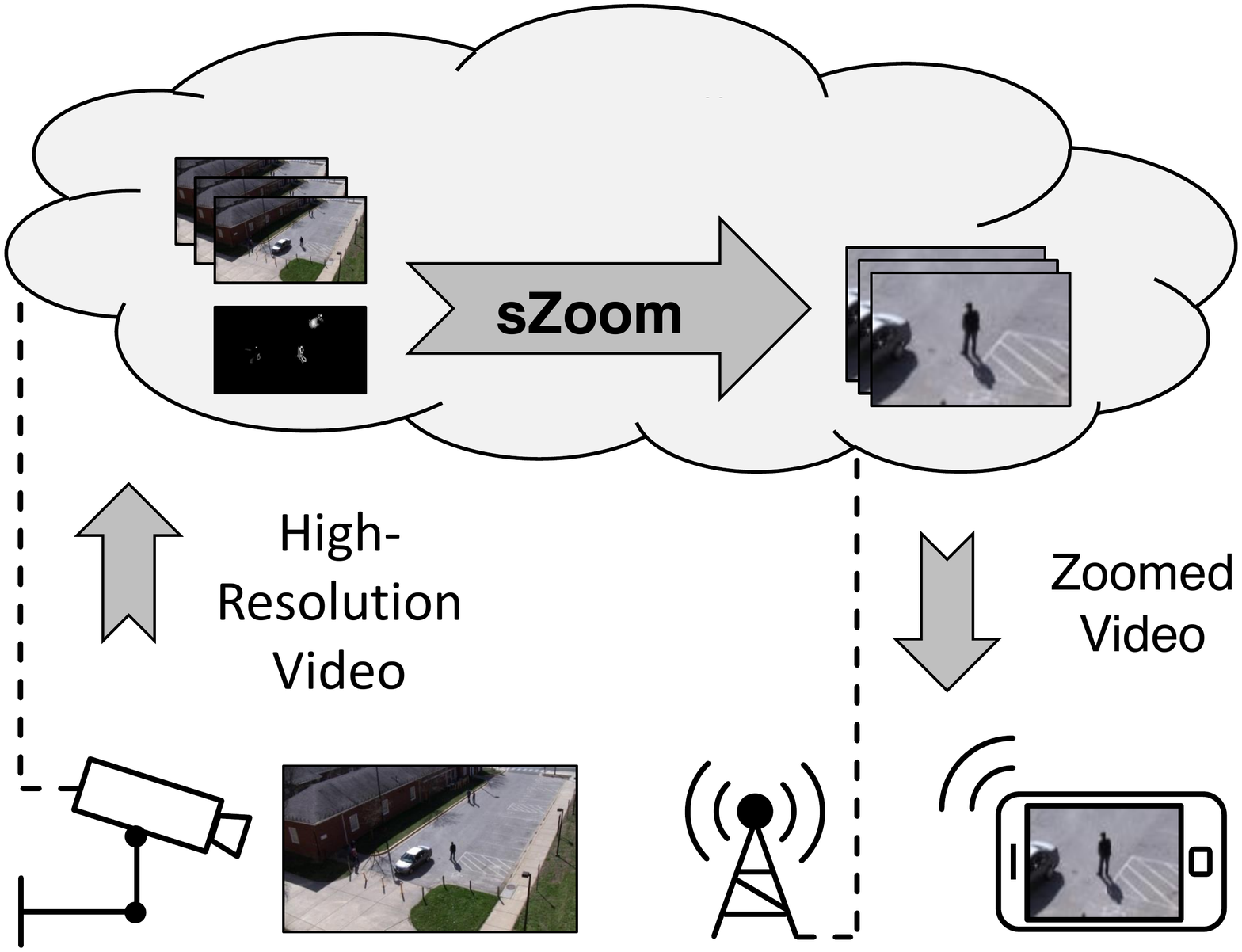}
\caption{The proposed sZoom framework. sZoom automatically zooms into the sensitive regions of the incoming surveillance video. In this way, the viewer gets a detailed view of the site on a small display device.}
\label{fig:cloud}
\end{figure}

In this paper, we propose  \textit{sZoom}, a novel automatic zoom framework for surveillance videos. 
This is shown in Figure~\ref{fig:cloud}. In the proposed method, we detect sensitive regions of a high resolution video and digitally zoom onto those regions. Sensitive regions are presented in more detail, while still providing to the operator the environmental context that is required for situation assessment. 

sZoom determines sensitivity of the regions in the video based on a set of semantic observations. The semantic observations are chosen according to the security threats at the surveillance site.  We aggregate these observations over a certain window of time to reduce noise. To ensure coverage, a decaying multi-variate Gaussian is employed to generate a penalty map that penalizes the regions selected in the previous cycle so that the other sensitive regions get priority in the near future. An iteratively  refined cubic spline function is used in tandem with target ROI tracking to obtain a consistent and natural zoom. The process of region selection and presentation is repeated for the entire duration of the surveillance.
The prototype of the framework is implemented for viewing a full HD video on a smartphone; with a case study of three semantic observations: motion, human body, and faces. 


The system performance is evaluated through three user studies with more than 78 participants and four full HD surveillance videos. In the first study, users were asked to rate their agreement with five statements about the appropriateness of the video for a surveillance task. The video produced by the proposed method received 46\% higher ratings. In the second user study, we measured the information gained by watching our processed video compared to a baseline system. The information gain was assessed by asking the participants questions about details in the videos after watching. On the average, 99\% more questions were answered correctly when watching the video created with sZoom. The results indicate that our proposed system provides a more useful view of the surveillance site.

The third experiment compares the proposed work with the preliminary version of the work that was published at as a short paper \cite{kuang2014real}. In this paper we improve the previous work with better sensitivity detection process,  reduced latency, and better coverage of moving targets. To deal with the motion of the target ROI, we iteratively adapt the cubic spline parameters based on tracking results.  
Results show that sZoom framework is more appropriate for surveillance task with 36\% higher ratings than the previous work.

The main contributions of this work are the following:

\begin{itemize}
\item We propose a novel method to automatically zoom into the surveillance videos which provides a more efficient presentation of the surveillance videos on a smaller display device. The proposed method improves the amount of information gained by security operators by 99\%.
\item We propose optimizations in the system that improve the accuracy by 42\% and reduce the processing time to 5\% of the original time making the sZoomed video more suitable for surveillance with 36\% higher ratings  in comparison to the previous work \cite{kuang2014real}.
\end{itemize}

The rest of the paper is organized as follows. Section \ref{SSA: Smart Surveillance Assistant} presents the proposed system and its components. Experimental results are provided in Section \ref{Experimental Results}. In Section \ref{Related Work}  we review and connect with the relevant literature. Finally, Section \ref{Conclusions} concludes the paper. 
\section{Proposed \MakeLowercase{s}Zoom Framework}\label{SSA: Smart Surveillance Assistant}
\begin{figure}
\centering
\includegraphics[width=0.8\linewidth]{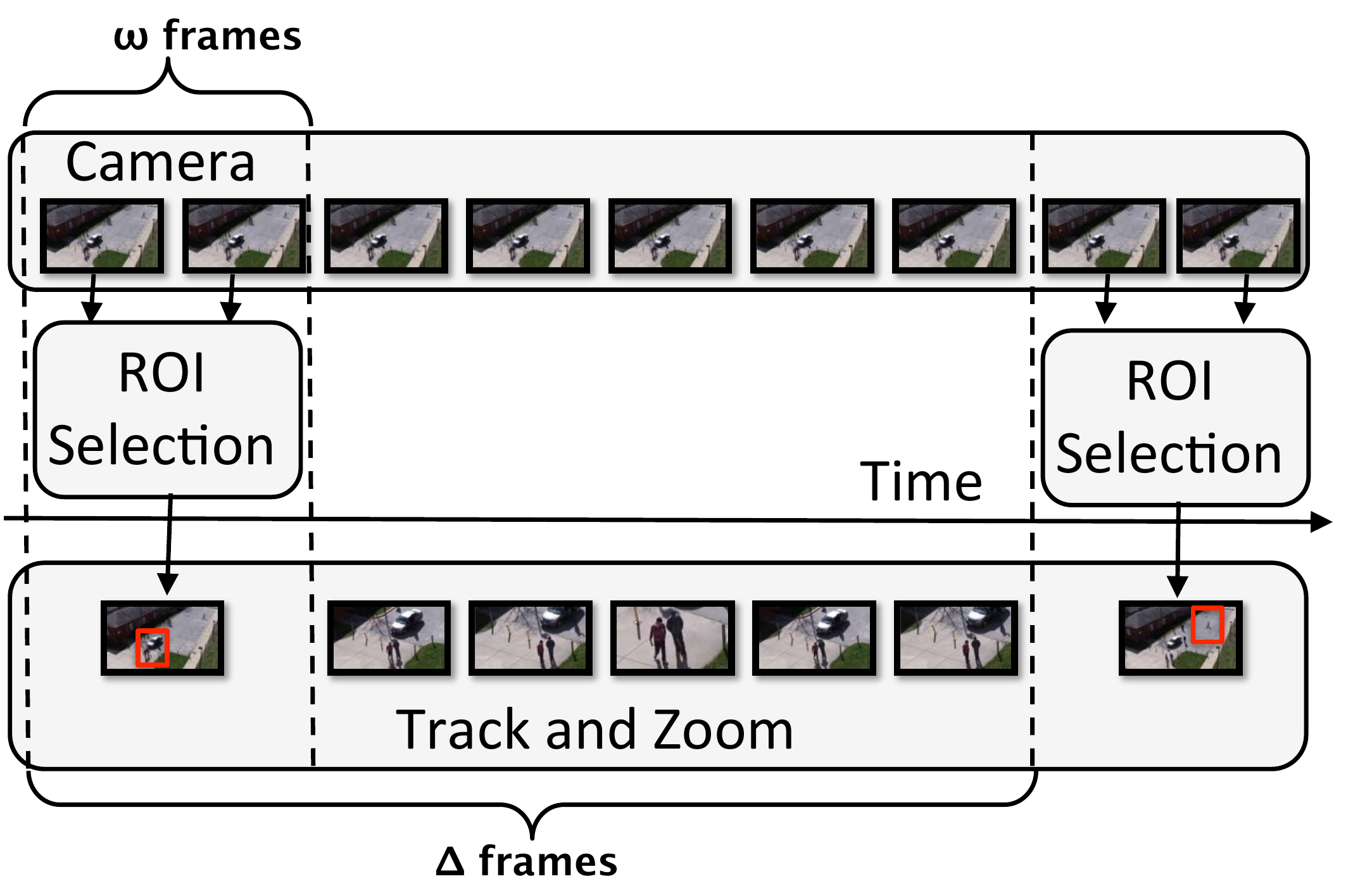}
\caption{
For every cycle of $\Delta$ frames, the system analyzes the first $\omega$ frames of the video with respect to importance from a surveillance perspective in order to select an ROI to zoom into.
While zooming, the system tracks and adjusts the ROI to compensate for object motion during the zoom operation.
}
\label{fig:overview-new}
\end{figure}

An overview of the method is shown in the Figure \ref{fig:overview-new}. 
Once every $\Delta$ frames, the method chooses a target region of interest (ROI) for the zoom operation. In order to minimize the system latency, only
 first $\omega$ frames are analyzed to predict the ROI for the whole zoom window ($\Delta$ frames). A tracking based reinforcement step is introduced to adapt for an ROI that contains moving targets. Note that it is important to focus on the same ROI for sufficient time to allow the operator to comprehend the situation.   
 
The initial ROI is chosen using multimodal fusion of persistent observations. Here a modality refers the semantic observations made by analysing the video. It is found that the non-persistent observations are generally due to noise. The persistence is calculated over $\omega$ frames, which is found different for different semantic observation in the experiments. The semantic observations depend on the threat definitions at the given surveillance site. An example of semantic observation is humans wandering in a restricted area.

For a surveillance task, coverage of the area is also important as there can be multiple threats simultaneously. Furthermore, a security operator requires contextual information to fully assess the activity happening in the zoomed ROI.  To ensure coverage, an online learned multi-variate Gaussian is employed. Finally, an iteratively refined cubic spline method is proposed to zoom onto the selected ROI. The iterative refinement is based on the updated parameters of the ROI past $\omega$ frames. The fused semantic observations are combined with user prior knowledge and the coverage function to obtain the final target ROI.  

In the remainder of this section, we discuss details of each component of the system.

\subsection{Threat Model}
The main purpose of the threat model is to identify regions of the video that are important and sensitive from a surveillance perspective and show those in more detail to the operator. In order to detect these regions, we need to make semantic observations.  For example, Hu et al. ~\cite{hu2004survey} show that the most relevant regions for surveillance are faces, human body, and regions belonging to moving objects in general. There is no fixed set of such observations that applies to all scenarios; it may vary across sites, and for the same site also over time. Therefore we have proposed a modular threat model that allows users to easily plug-in new semantic observations. 

The multiple semantic observations are fused using convex combination, where the weight of each observation is calculated according to the accuracy of the corresponding semantic observation. If each semantic observation is made with a corresponding detector, the sensitivity of a pixel is calculated as:

\begin{equation}
\mathcal{I}_{ij}^o (t) = \sum_{k=1}^{n}{c_kO_{ij}^k}(t)
\label{eq:combined}
\end{equation} 

where $O_{ij}^k$ is  the $k^{th}$ semantic observation for the pixel with $(i,j)$ index of the image and $c_k$ is the conformal coefficient of the $k^{th}$ semantic observation that depends on the accuracy of the corresponding detector. In this equation, the coefficient $c$ accounts for the inherent  uncertainty of the detector. The conformal coefficients are chosen to reflect our confidence in each semantic observation. We choose convex combination to fuse multiple observations as it reinforces areas with multiple threats.     

During the case study (Section \ref{sec:case-study}), we found that with the current analysis methods, semantic observations are not entirely accurate. The observations are noisy at times due to limitations of the detectors. To minimize the effect of noise, we determine the persistence of the observations over a window of $\omega$ frames: 

\begin{equation}
O_{ij}^k(t) = \frac{1}{\omega}\sum_{f=t-\omega+1}^{t}I_{ij}^k(f)
\label{eq:accumulation}
\end{equation}

where $I_{ij}^k(f)$ is observation in an individual frame. Note that this accumulation step adds a latency of $\omega$ frames to the system. However, in the experiments it is found that $\omega <<\Delta$. Hence, we get an improved prediction at the cost of small latency. In the previous work \cite{kuang2014real}, the latency was a full zoom window, i.e., $\Delta$ frames. We are able to reduce it to $\omega$ frames because of the ROI tracking and corresponding iterative refinement of the target ROI during the zoom operation (Section \ref{Natural and Consistent Zoom}).

\subsection{Fair Coverage}
At any given time, only one region can be selected as target ROI. If we only rely on the threat model described above, we may end up selecting the same ROI as target in each zoom window. This is not fair to slightly less important regions which also need operator's attention. Furthermore, the other less important regions ensure contextual knowledge to the operator. A similar situation occurs while scheduling processes in operating systems. There, a low priority task may never get scheduled. To build a completely fair scheduler, the priority of a process is updated (reduced) after it is scheduled for execution.   

We also take the similar approach and penalize the regions that have been selected in the recent past.
For each selected ROI, the system adds a multivariate Gaussian function $\mathcal{G}(\mu_t, \sigma^x_t, \sigma^y_t)$ to the penalty map $\mathcal{P}_t$.
Its mean $\mu_t$ is the center of the ROI and the variances $\sigma^x_t$ and $\sigma^y_t$ are chosen according to its width and height. The Gaussian function models the probability of presence of the object within the rectangular ROI. 
The penalty values of the map decay over time, so that previously chosen ROIs can be chosen again after a certain time.
This decay is implemented by multiplying the penalty map by a factor $\alpha \in [0, 1]$ each time a new ROI is selected.
The penalty map is calculated at frame $t$ as:
\begin{equation}
\mathcal{P}_t= \alpha*\mathcal{P}_{t-\Delta} + \mathcal{G}(\mu_{t}, \sigma^x_{t},\sigma^y_{t})
\label{eq:penalty}
\end{equation}
where $t-\Delta$ is the frame index at the time the ROI was selected in the previous cycle.
Since the ROI chosen in the previous cycle is tracked continuously for the duration of $\Delta$ frames, i.e., an entire cycle, the parameters of the Gaussian function are based on the position and size of the ROI in the \emph{last} frame.
In our experiments, we chose $\sigma^x_t$ and $\sigma^y_t$ to be half the width and height of the ROI, respectively. We penalize the current sensitivity with this penalty map.

Another limitation of the threat model is that it cannot differentiate usual motion, such as traffic on the road, from the activities in the parking lot. To address this, sZoom takes explicit input from the user where they clearly specify the regions of the camera view that are not relevant. 
The user provides input in terms of a binary map $\mathcal{U}$ of the same size as the image.
A 1 in $\mathcal{U}$ marks a pixel as belonging to a relevant region, and 0 indicates irrelevant ones.

In summary, the penalty map and user input ensure that:
\begin{enumerate}
\item The regions such as an adjacent road with traffic that contain motion which is not relevant to the surveillance are not selected. 
\item Most regions of the scene get selected to improve coverage. Note that a round robin kind of scheduling would not work in this case as the ROIs are volatile. 
\end{enumerate}

\begin{figure}
\centering
\subfigure[]
{
\includegraphics[width = 0.45\linewidth]{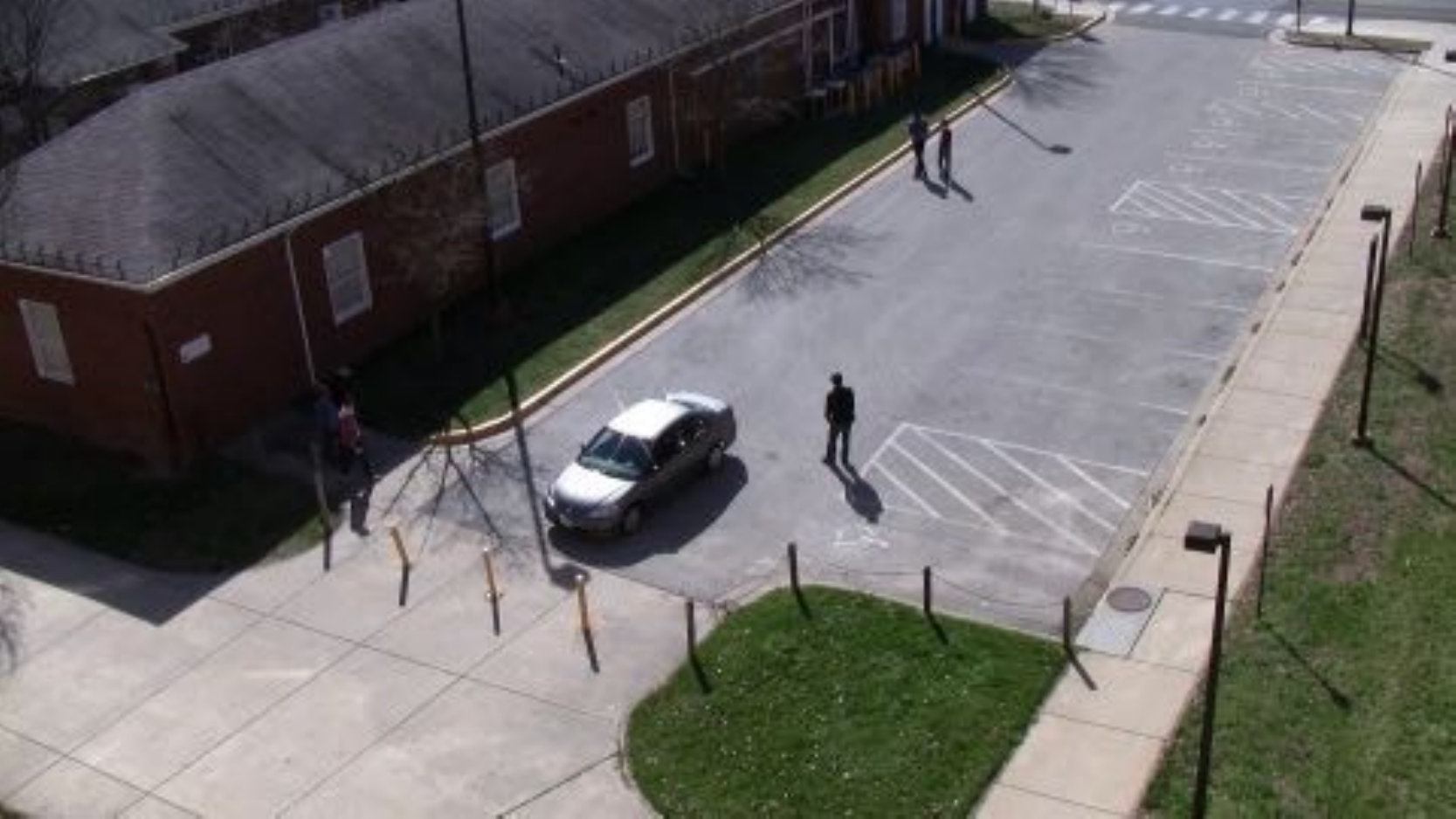}\medskip 
\label{fig:penaltyoriginal}
}
\subfigure[]
{
\includegraphics[width = 0.45\linewidth]{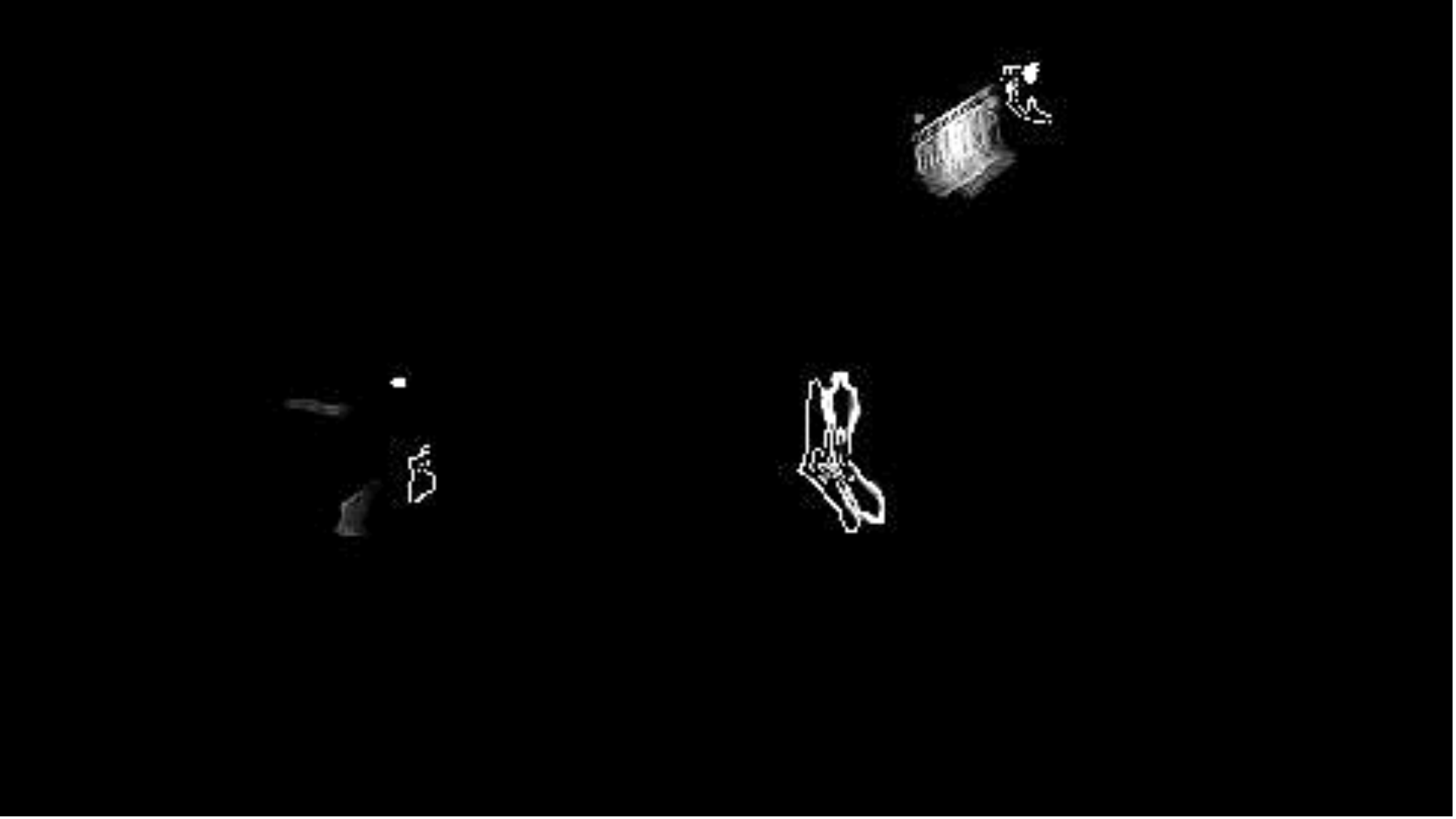}\medskip
\label{fig:penaltymask}
}
\subfigure[]
{
\includegraphics[width = 0.45\linewidth]{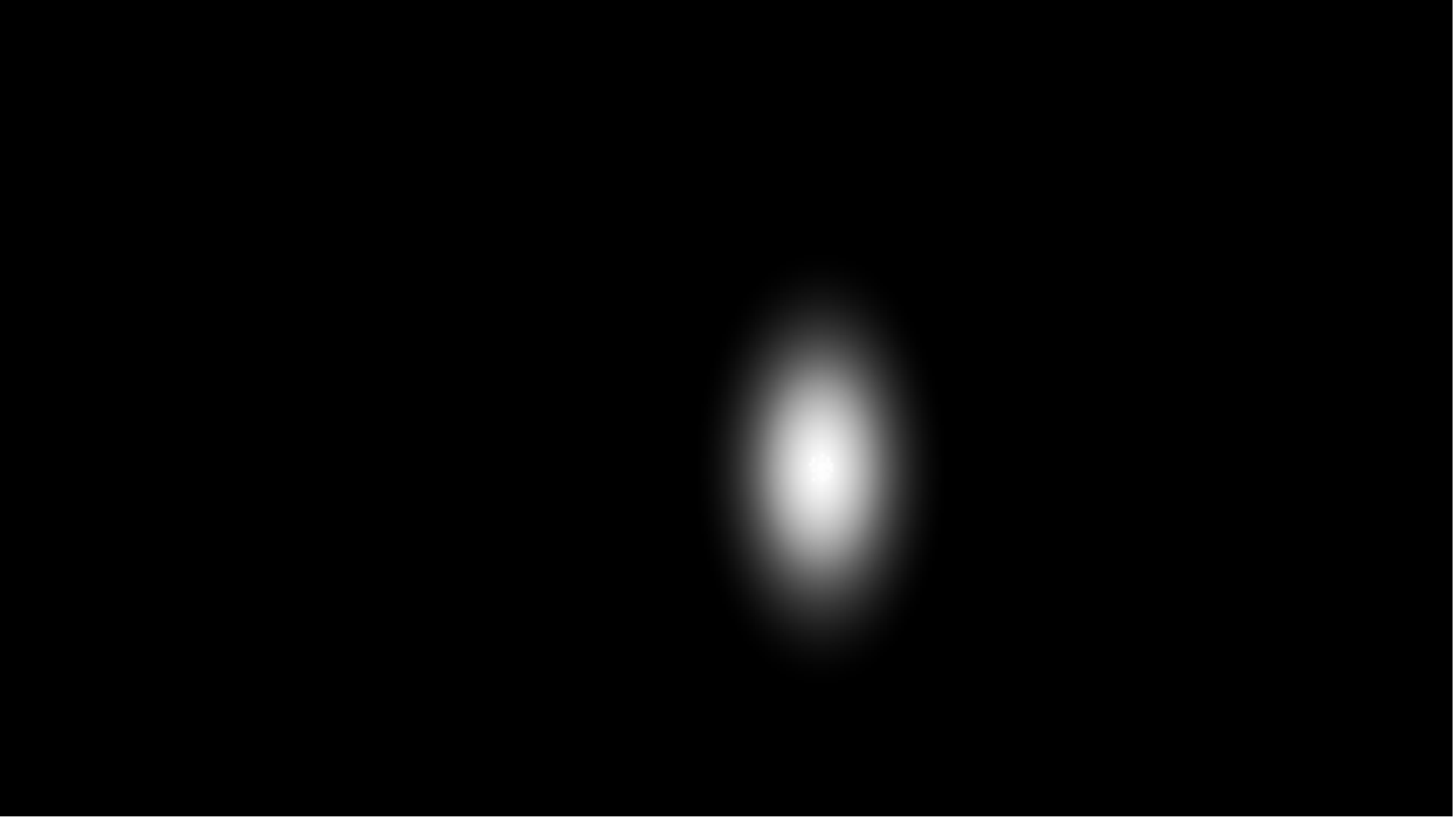}\medskip
\label{fig:penaltypenalty}
}
\subfigure[]
{
\includegraphics[width = 0.45\linewidth]{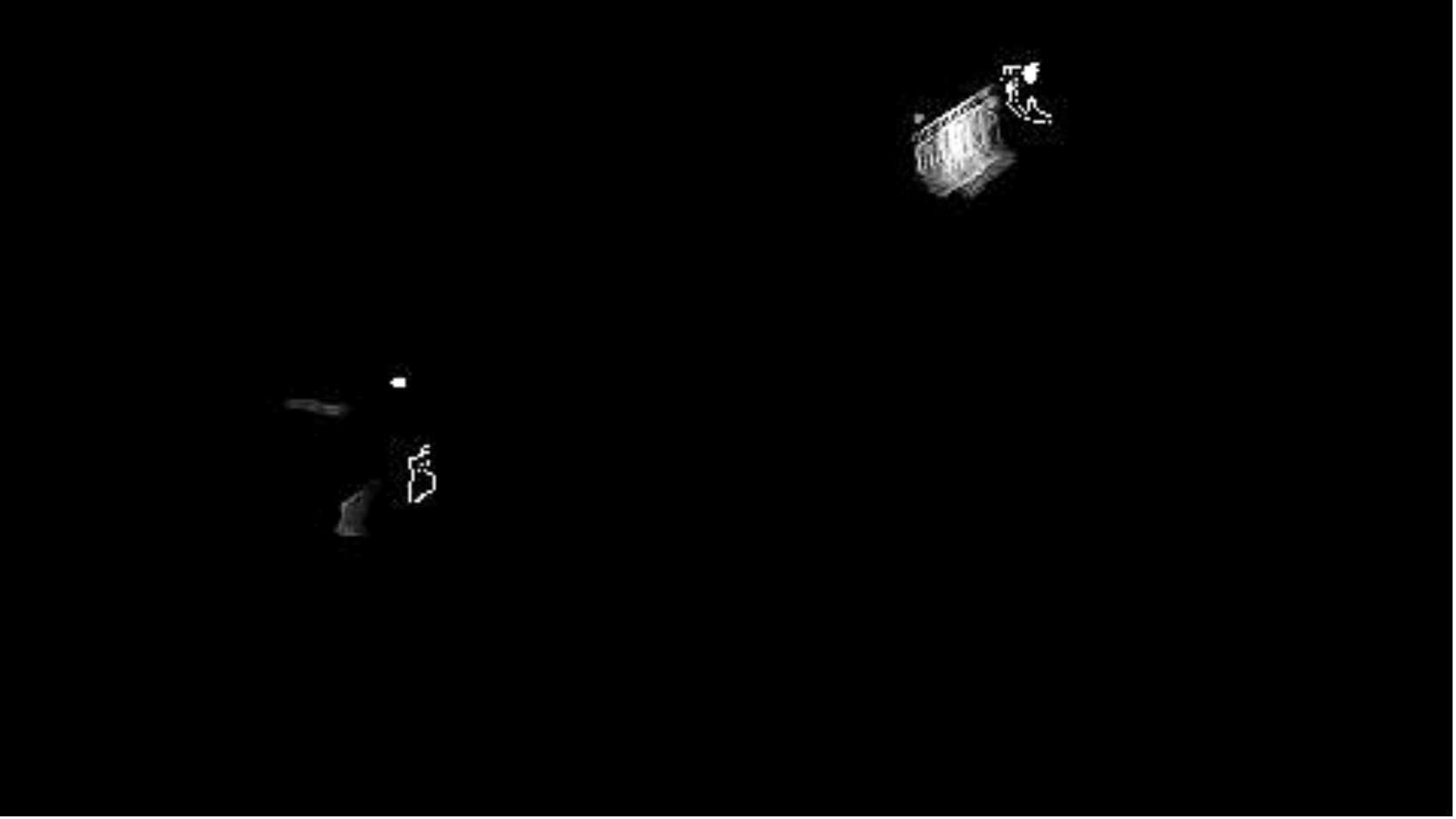}\medskip 
\label{fig:penaltydecision}
}
\caption{
An example of the penalty mechanism.
The current frame of the video is shown in (a).
The fused semantic observation $\mathcal{I}_{ij}^o (t)$  is shown in (b).
The penalty map $\mathcal{P}(t)$ (c) penalizes the previously selected ROI.
As a result, the decision map $\mathcal{D}_t$ (d) no longer contains the sensitivity of the person in the center. The test video (a) is taken from  VIRAT dataset \cite{oh2011large}.
}
\label{fig:penalty}
\end{figure}

While the user input provides binary information on the importance of regions, the penalty map specifies how recently a particular region of the video was selected for zooming. With this information, a decision map $\mathcal{D}_t$ is calculated by fusing sensitivity, user input, and penalty map as follows:
\begin{equation}
\mathcal{D}_t= (1-\mathcal{P}_t) \cdot (\mathcal{U} \cdot \mathcal{I}_t)
\end{equation} 

where $\mathcal{I}_t$ is the sensitivity of entire image obtained with Equation \ref{eq:combined}. Figure \ref{fig:penalty} shows an example of how a previously selected ROI is penalized. Fig. \ref{fig:penaltyoriginal}  shows the original image, and its corresponding sensitivity map is shown in Fig. \ref{fig:penaltymask}. The person in the center of the frame has been selected as the area to zoom into in the previous cycle.
In the current cycle, the position of the person is penalized by the penalty map shown in Fig. \ref{fig:penaltypenalty}.
As a result, the man will not be selected again in the decision map in Fig. \ref{fig:penaltydecision}.

The decision map contains values between 0 and 1. A threshold is applied to the decision map to convert it into a binary image.
Contours are retrieved from the resulting binary image using the border following approach given in \cite{suzuki1985topological}.
Nearby contours are merged together while too small contours are discarded.
The bounding boxes around the remaining contours constitute the candidate ROIs.
They are ranked according to the sum of sensitivity over the ROI in the original decision map $\mathcal{D}$.
The rectangle with the highest total sensitivity is then chosen as the target ROI.  

The output of the ROI selection step is the rectangle which is most likely to contain objects of interest.
We employ a modified mean shift tracking algorithm to track objects in the RGB frame sequence during the entire cycle~\cite{comaniciu2000real}.
A constraint that the ROI size remains invariant is added to the tracking method.
The assumption that the size of the tracked object remains constant is true for small periods (e.g. 5 seconds in experiments).
With the tracking mechanism in place, new coordinates for the target ROI are obtained in each frame.    

\subsection{Natural and Consistent Zoom}\label{Natural and Consistent Zoom}
Zoom does not exist in the real world. Therefore, a sudden full speed zoom would cause discomfort to the viewer. To mitigate this discomfort, we need to have low zoom speed at the beginning and the end. The speed should gradually increase and decrease in the middle. We found that a cubic spline function can be customized to meet these requirements.   

The aspect ratio of the selected ROI is arbitrary and generally differs from the aspect ratio of the target resolution. The width $w$ and height $h$ of the ROI thus need to be adjusted to match the target aspect ratio. If the aspect ratio of the ROI $\phi = w/h$ is less than the target aspect ratio $\phi'$, $w$ is increased on both sides of the rectangle. If $\phi>\phi'$, $h$ is increased accordingly.

The process of zooming involves scaling the input frame, panning, and cropping the target ROI. Panning is required to keep the ROI in the center. It is implemented by image translation. The zoom operation starts and ends with a fully zoomed out view, i.e., an input video scaled down to the target ROI size. In between, the system zooms into the ROI, stays there for a while and then zooms out again. Hence, the zoom parameters in each frame are the horizontal and vertical position as well as the width and the height of the rectangular window that is currently presented to the viewer.

In each frame, these parameters are obtained using cubic Hermite spline interpolation to achieve a natural zoom effect.
Each zoom parameter is interpolated individually.
If $A_0$ is the start value of a parameter and $A_1$ is the value at the end, an intermediate value of the parameter $A_t$ for frame $t$ is obtained as follows:   
\begin{equation}
A_t = A_0(2f_t^3 - 3f_t^2 +1) + A_1(-2f_t^3 + 3f_t^2)
\end{equation}
where $f_t$ is a value between 0 and 1 that denotes the time that has passed between the first frame ($t_0$) and the last frame ($t_1$) of the zoom operation.
It is defined as $f_t = (t-t_0)/(t_1-t_0)$. In the interpolation, we also enforce a zero tangent, i.e., no pan and zoom motion, at the start and the end points. Note that while the initial parameter values $A_0$ remain constant, the final values $A_1$ are updated after each iteration according to the results of tracking. Because the location of the target ROI may change in each frame due to tracking, using the values directly may result in a jerky zoom operation.
To avoid this artifact, a median filter is applied to the calculated zoom parameters in order to ensure a smooth zooming operation. %

We call the final zoom operation \emph{AB} zooming. In this scheme, the video has a fixed zoom for A\% of the cycle followed by a continuous zoom over B\% of the frames. The fixed zoom can be no zoom, i.e. a scaled down version of the input video, or a fully zoomed video, i.e., the final ROI. For the given dataset, we found A = 20 and B = 30 to be appropriate.  During the first 20\% of the cycle of $\Delta$ frames, the system presents the entire input video scaled down to the target size. Afterwards, the system zooms onto the chosen ROI over the course of $30\%$ of the cycle. The view remains zoomed in for $20\%$ of the cycle and then zooms again over the course of the last $30\%$ of the cycle. 

The sZoom processes the high resolution surveillance video streams in the cloud and adapts them for mobile devices. It produces a low resolution stream to be viewed on hand-held devices with small displays, where the exact size of the target video depends on the physical dimensions of the screen. We argue that a resolution of 72 PPI is sufficient for surveillance purposes.
For a full HD input and typical sizes of the hand-held devices ($\approx 5$ inches display size), the size of the output video is thus 5 to 6 times smaller than the input size. 
  

\subsection{Case Study: Motion, Humans, and Faces as Semantic Observations}\label{sec:case-study}

This section discusses a case study with three semantic observations: motion, humans, and faces. In the literature these observations have been found important across most surveillance sites \cite{hu2004survey}. The method can be scaled easily to include more threats simply by adding the observations and calculating the accuracy of the corresponding detector. We demonstrate the effectiveness of the proposed framework with baseline detectors to simulate the worst case scenario. We choose the detectors that are not so accurate, but run in real-time. Any better detector will further improve the sZoom performance. 
\subsubsection{Semantic Observation 1: Motion}
Motion is observed in the regions of the video that are occupied by moving objects. There exist various methods to detect motion in a video such as temporal differencing, optical flow, and background modeling \cite{hu2004survey}.
In our work we choose background model based algorithm by Zivkovic et al. \cite{zivkovic2006efficient} that uses customized MoG for each pixel, providing better adaptability to scenes with varying illumination. 
The entire process of foreground detection is is follows: 
\begin{itemize}
\item Resize the image to a smaller resolution to reduce the processing time. The resize factor is obtained through experiments reported in Section \ref{Timing Analysis}.
\item Determine the foreground mask using~\cite{zivkovic2006efficient}.
\item Apply dilation and erosion to reduce noise in the foreground mask. A default kernel of size $3\times 3$ is used for both erosion and dilation. 
\item Find the contours in the smaller binary image and fill rectangles encompassing these contours. 
\item Resize the rectangles according to the original size of the image. These rectangles contain the foreground objects. 
\item The final motion observation $I^{m}(t)$ is calculated as the union of the areas of all rectangles.
\end{itemize}

\begin{figure}
\centering
\subfigure[]
{
\includegraphics[width = 0.45\linewidth]{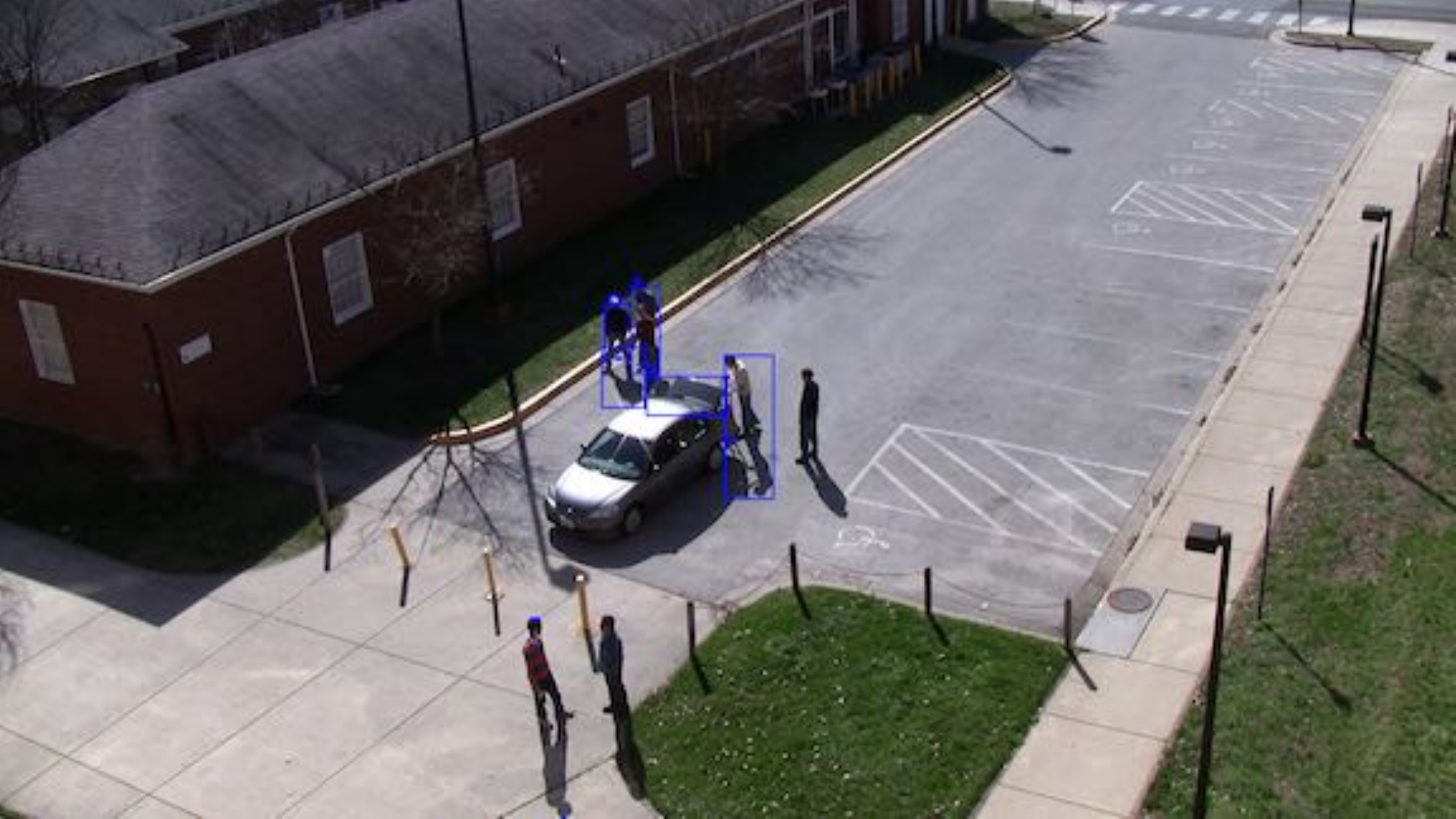}\medskip 
\label{fig:foreground1}
}
\subfigure[]
{
\includegraphics[width = 0.45\linewidth]{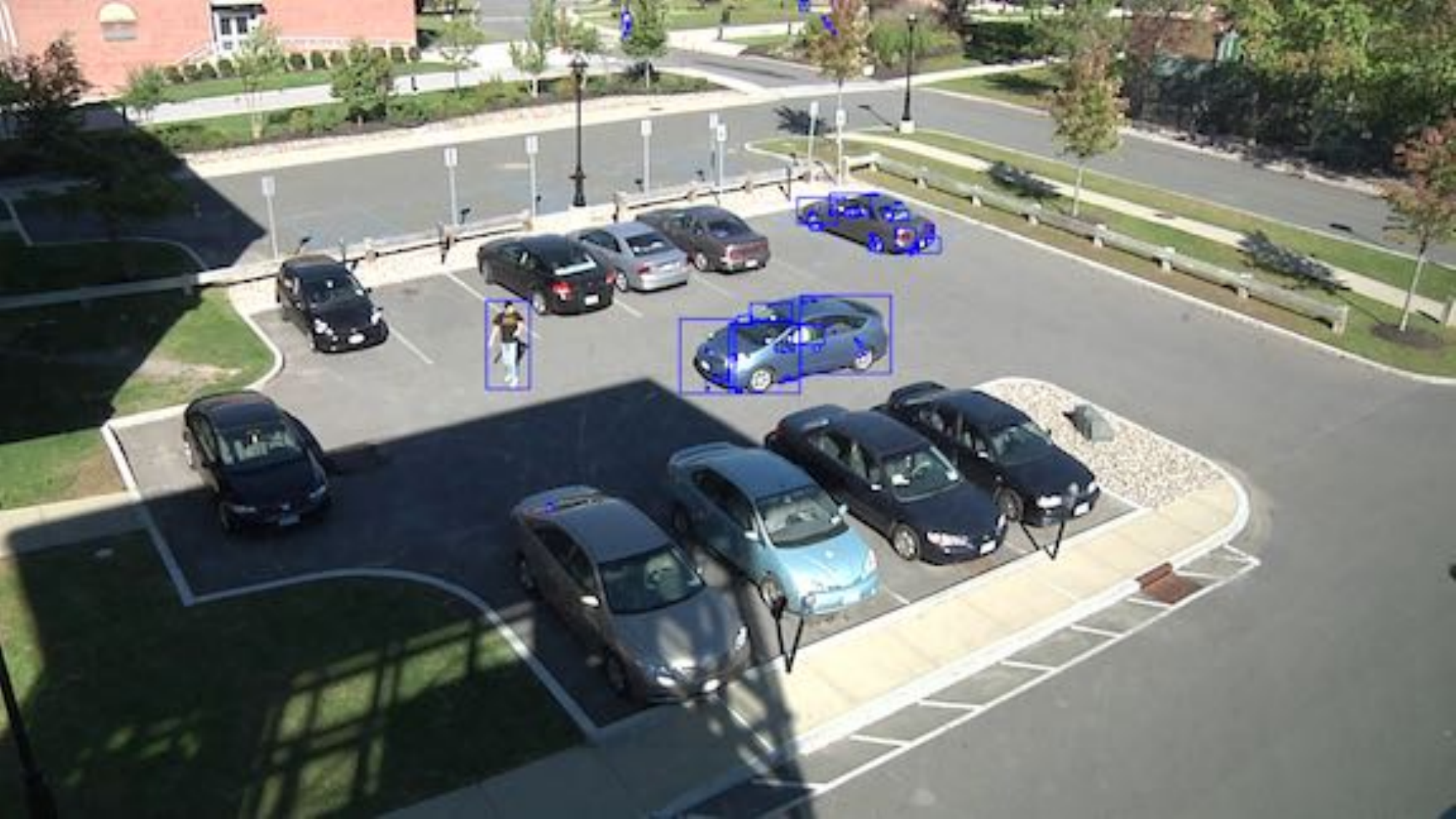}\medskip
\label{fig:foreground2}
}
\subfigure[]
{
\includegraphics[width = 0.45\linewidth]{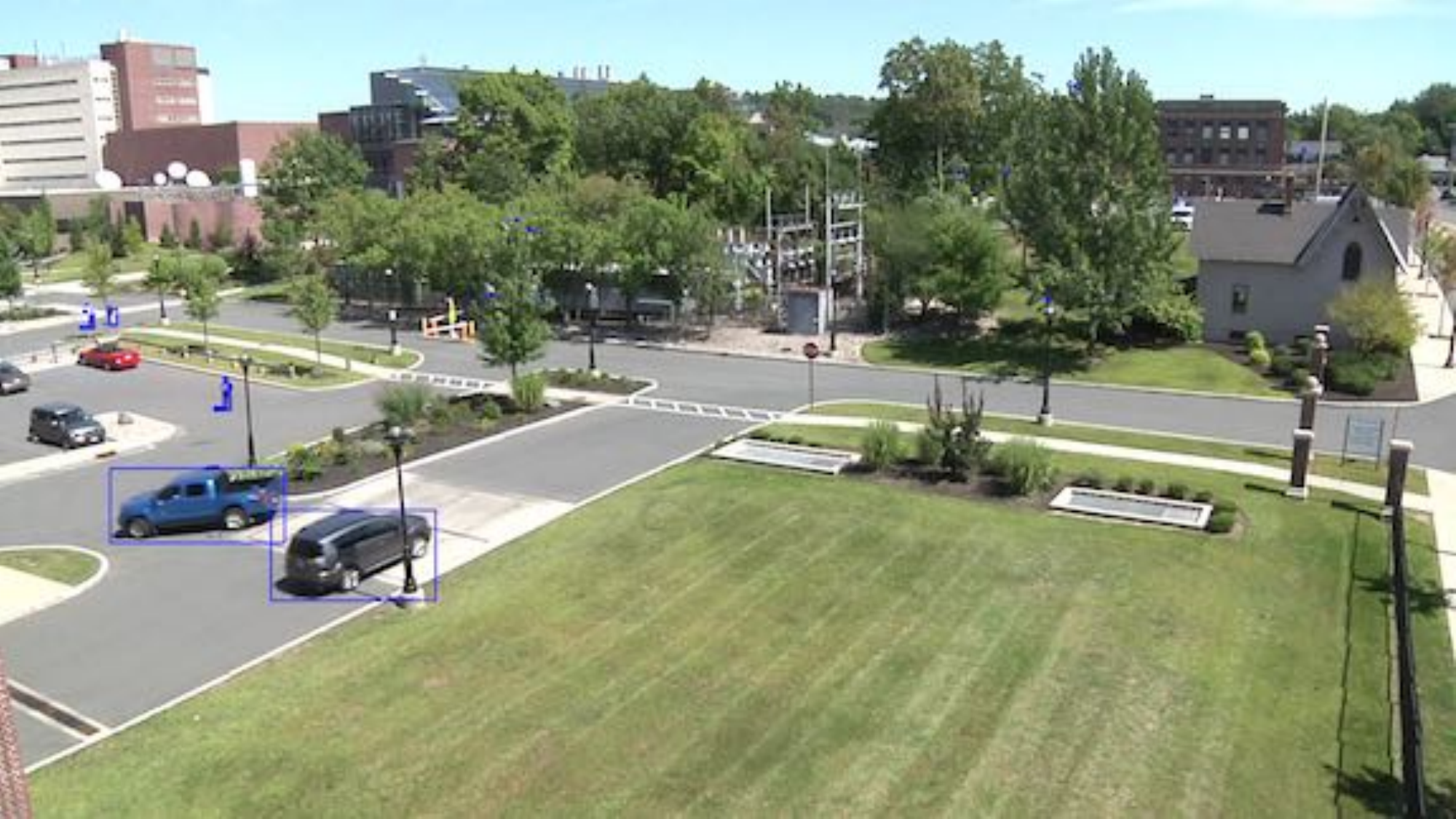}\medskip
\label{fig:foreground3}
}
\subfigure[]
{
\includegraphics[width = 0.45\linewidth]{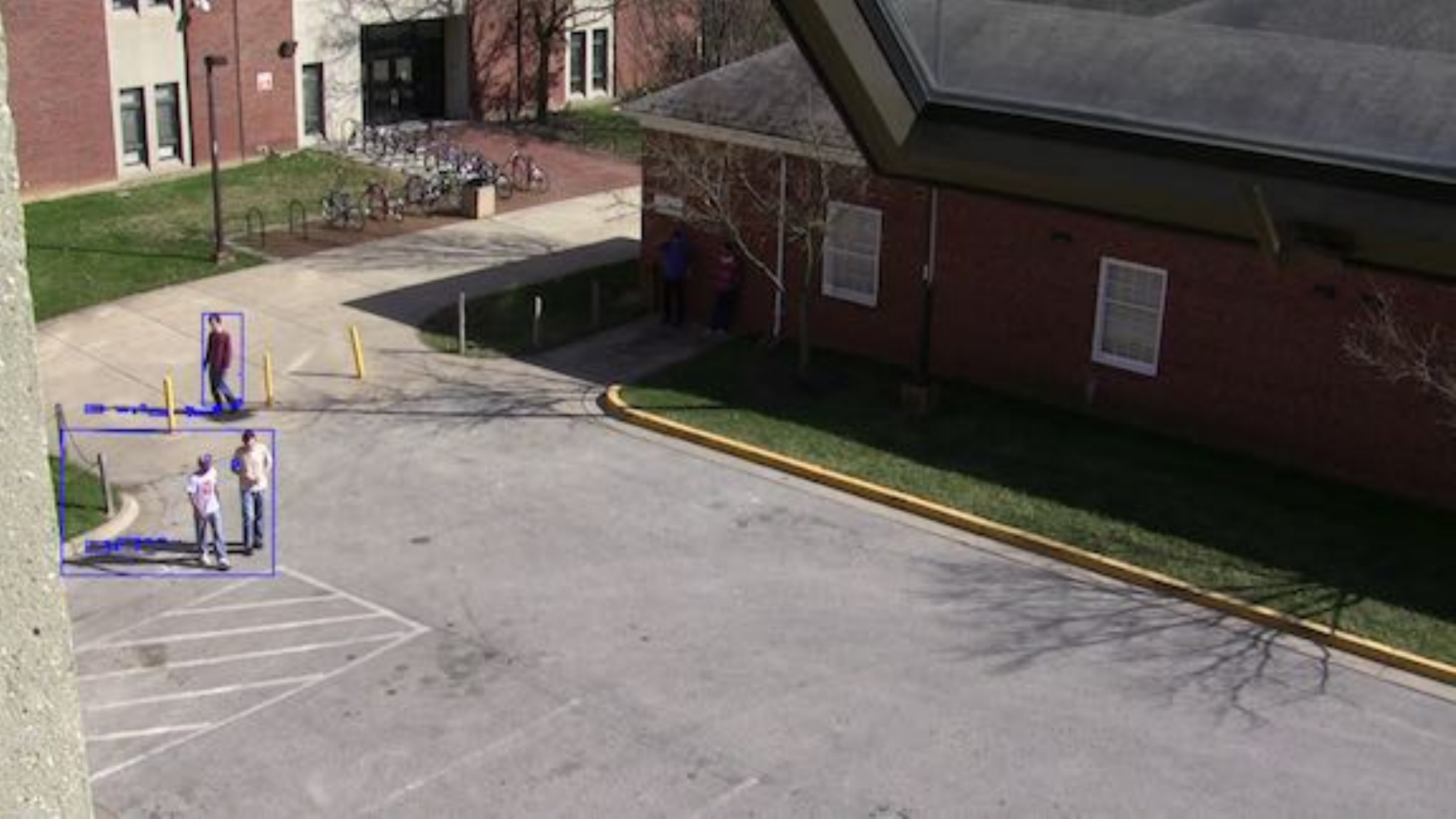}\medskip 
\label{fig:foreground4}
}
\caption{Example images with foreground detection results shown as blue rectangles. The algorithm is able to detect moving cars in (b) and (c) and humans in all four figures. However, it fails to detect humans that are standing still in (a) and standing in the shadow of the house in (d). The original images were taken from  \cite{oh2011large}.}
\label{fig:foreground}
\end{figure}

Figure \ref{fig:foreground} shows illustrative results of the detector as blue rectangles around the moving objects. It can be seen that the algorithm fails to detect humans who are not moving (Figure \ref{fig:foreground1}) or standing in the shadow (Figure \ref{fig:foreground4}). In surveillance, however, humans need to be detected even if they are not moving. This is accomplished by the second semantic observation. 


\subsubsection{Semantic Observation 2: Humans}
With the evolution of deep learning techniques, there are various methods that detect humans very accurately, such as YOLO object detector \cite{redmon2016you}. However, we choose Histogram of oriented Gradients (HoG) descriptor to represent human bodies~\cite{dalal2005histograms} to demonstrate the framework as it works in almost real-time. The method divides the image into small cells and calculates a 1D histogram of gradient directions for each cell~\cite{poppe2010survey}. The cell size is chosen as $8\times8$ pixels. The HoG features are used as input to a support vector machine based classifier for body detection.



\begin{figure}
\centering
\subfigure[]
{
\includegraphics[width = 0.45\linewidth]{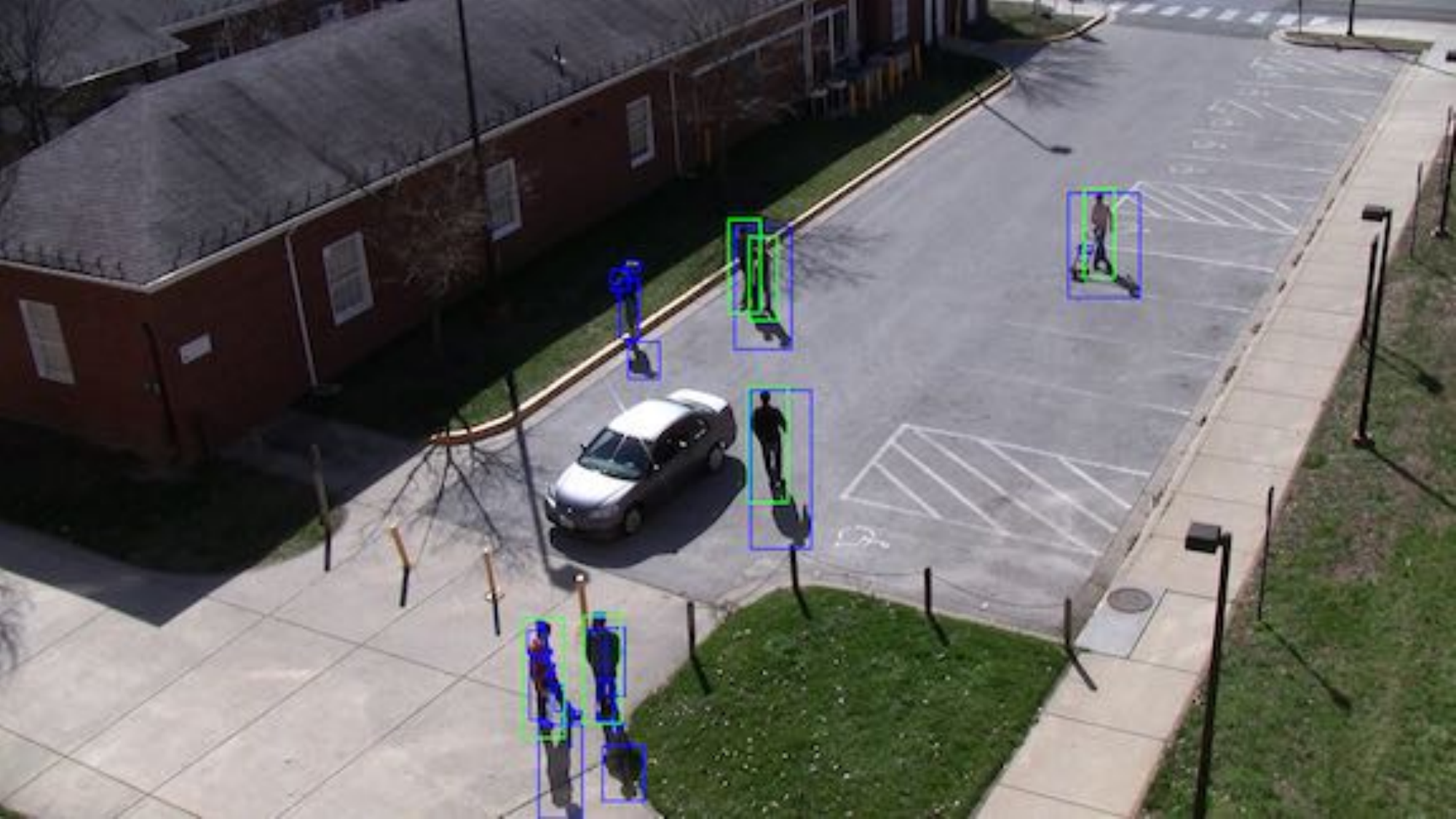}\medskip 
\label{fig:body1}
}
\subfigure[]
{
\includegraphics[width = 0.45\linewidth]{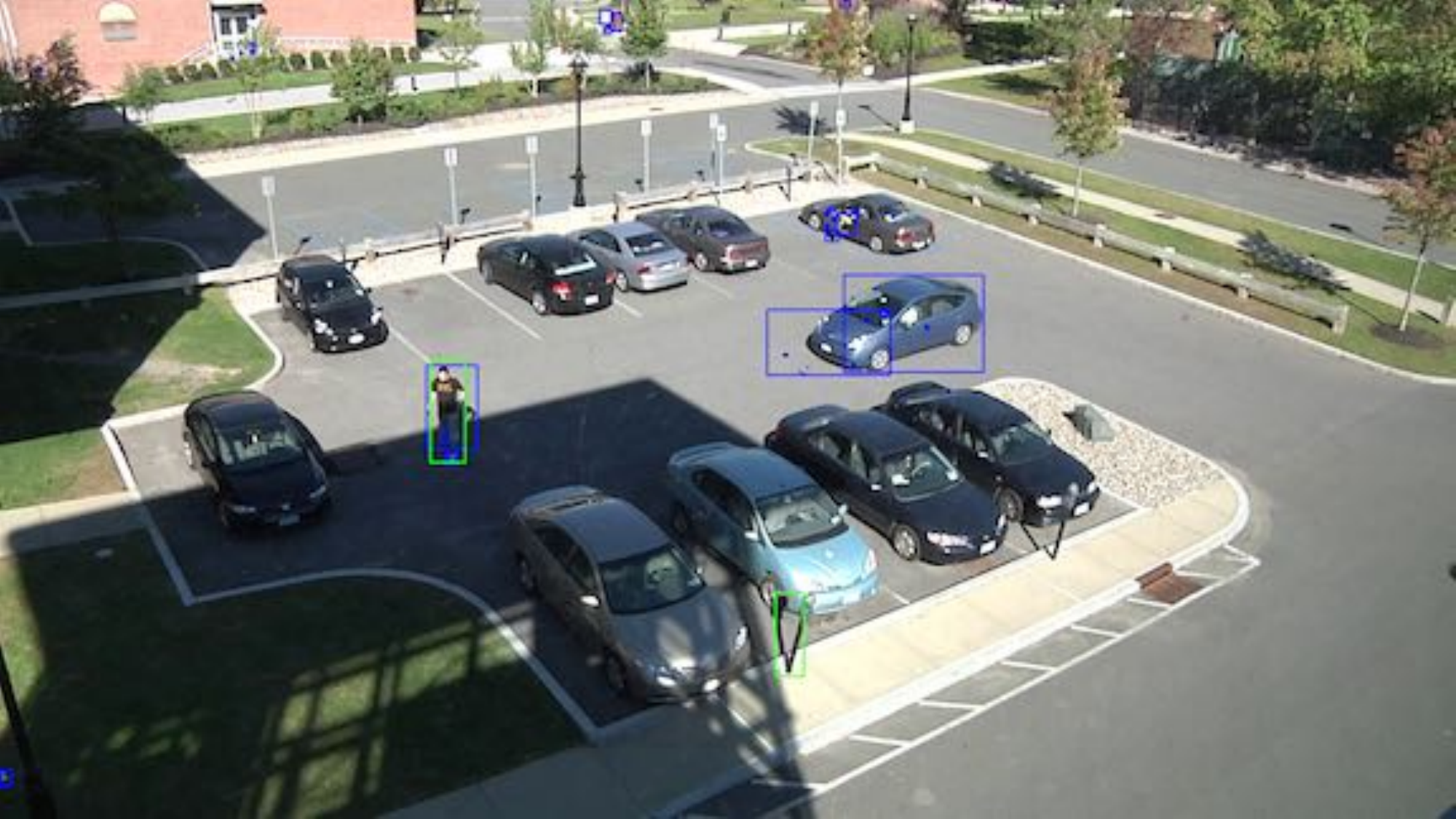}\medskip
\label{fig:body2}
}
\subfigure[]
{
\includegraphics[width = 0.45\linewidth]{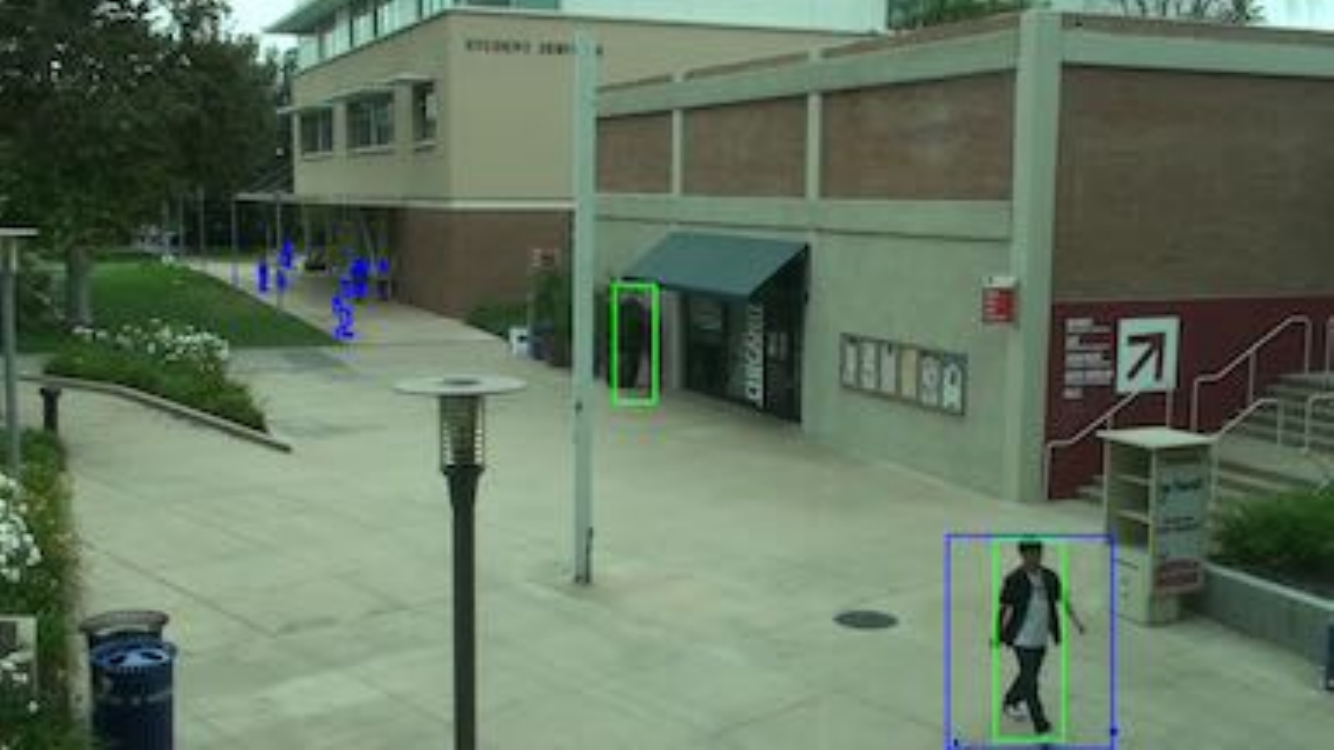}\medskip
\label{fig:body3}
}
\subfigure[]
{
\includegraphics[width = 0.45\linewidth]{E_119.pdf}\medskip 
\label{fig:body4}
}
\caption{Results of the human body detector shown as green rectangles. It can be seen that the body detection algorithm is able to detect humans that are standing still (a) and humans standing in the shadow (d). The detector is also able to differentiate between humans and other moving objects. The original images were taken from  \cite{oh2011large}.}
\label{fig:body}
\end{figure}

The algorithm yields a binary image $I^{h}(t)$ which has a value of 1 in the areas occupied by a human body, and 0 otherwise.
Exemplary body detection results are shown in Figure~\ref{fig:body}.
As opposed to the foreground detector, it can be seen in Figure \ref{fig:body1} that the body detector is able to even detect the humans who are not moving at the moment. Also, in Figure \ref{fig:body4}, the algorithm successfully detects the people standing in the shadow of the house.
The system is thus able to differentiate between humans and other moving objects.
This allows the system to assign higher weights to human bodies in the sensitivity map in order to accommodate the higher importance of humans in surveillance.

\subsubsection{Semantic Observation 3: Faces}
There has been numerous recent works on face detection that achieve high accuracy, such as OpenFace tool that uses deep learning for face detection \cite{baltruvsaitis2016openface}. Again, in our system we use a baseline Haar feature-based cascade classifiers proposed by Viola and Jones to detect faces~\cite{viola2001rapid}. Any advanced face detector will further improve the performance the proposed framework. It is a real-time method that is based on the three main ideas: calculation of an integral image, machine learning with AdaBoost, and the attentional cascade structure. The result of the face observation is a binary image $I^{f}(t)$ with 1's representing face regions and 0's otherwise.  

\subsubsection{Combined Sensitivity Map}

Neither of the three detectors described above gives perfect results.
We observed that false positives and false negatives only appear sporadically while the correct detections are persistent.
We make use of this observation by accumulating the detection results over $\omega$ frames according to Equation \ref{eq:accumulation}. 
%
Let the accumulated sensitivity maps for the motion, humans and faces be $O^m$, $O^h$, and $O^f$, respectively.  

For the combination of the results of the three detectors, we consider their respective accuracy.
Generally, for a given surveillance scenario, each detectors has a different accuracy. For example, if the camera is far from the monitored area and capturing a wide view, the accuracy of the face detector is low, whereas the motion detector works well in most scenarios. Therefore, Equation \ref{eq:combined} is used to combine the three observations as follows:   

\begin{equation}
\mathcal{I}^o = c_{m}*O^m + c_{h}*O^h + c_{f}*O^f
\label{eq:sensitivity}
\end{equation} 
The coefficients $c$ are chosen to be proportional to the accuracy of the corresponding detector. By choosing appropriate weights in Equation \ref{eq:sensitivity}, the system gives more importance to the accurate detectors, improving the system performance. Also note that the face of a moving person is detected by all three detectors. As a consequence, the facial region gets the highest sensitivity value, because it appears in Equation \ref{eq:sensitivity} three times. Similarly, a moving human body gets detected by the body and the foreground detectors. It thus appears twice in Equation \ref{eq:sensitivity}, giving it a higher sensitivity than general foreground objects but a lower sensitivity than faces.

\section{Experimental Results}\label{Experimental Results}

The experimental results consist of three objective evaluations and three user studies.
In the first objective experiment, we determine the optimal size of the accumulation window $\omega$.
In the second, we analyze the timing performance of the three employed detectors and determine the optimal image resolution for each of them.
The goal of this experiment is to reduce the processing time without affecting the accuracy significantly.
The third experiment compares the accuracy and the timing performance of the proposed system with the state of the art \cite{kuang2014real}.
Lastly, we assess the effectiveness of the threat model, coverage, and overall zoom operation through three user studies.
In the first study, the participants are asked to rate the quality of the produced videos according to five criteria of appropriateness for surveillance.
In the second study, we ask the participants about details of the content of the videos in order to evaluate whether or not our system facilitates understanding the events in the scene. The third user study makes a qualitative comparison of the proposed automatic zoom framework with the state of the art.  

\subsection{Selection of Accumulation Window Size}\label{Selection of Accumulation Window Size}
\begin{figure*}
\centering
\subfigure[Face]
{
\includegraphics[width = 0.5\linewidth]{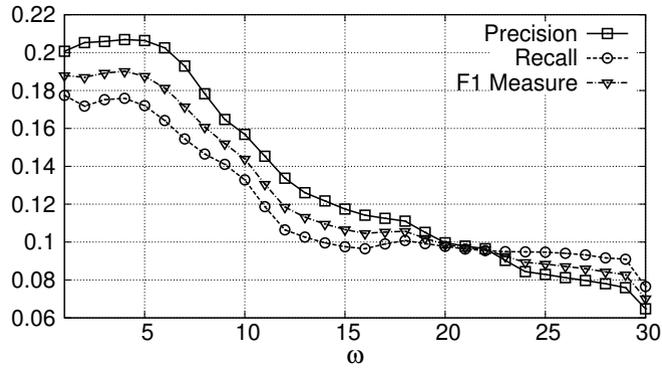}\medskip 
\label{fig:face-omega1}
}
\subfigure[Human]
{
\includegraphics[width = 0.5\linewidth]{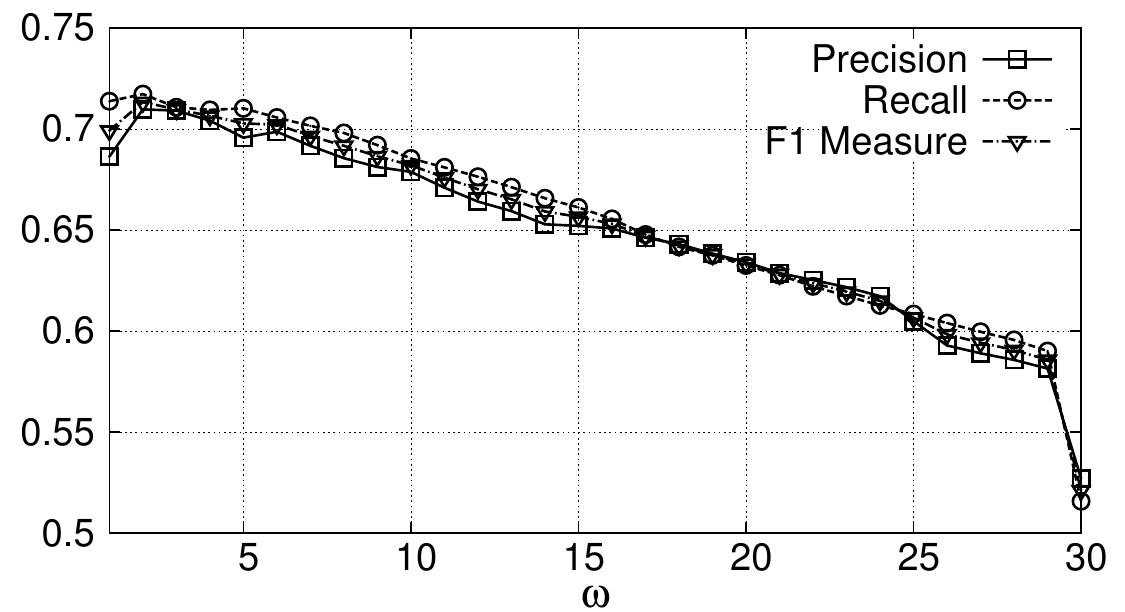}\medskip
\label{fig:face-omega2}
}
\subfigure[Motion]
{
\includegraphics[width = 0.5\linewidth]{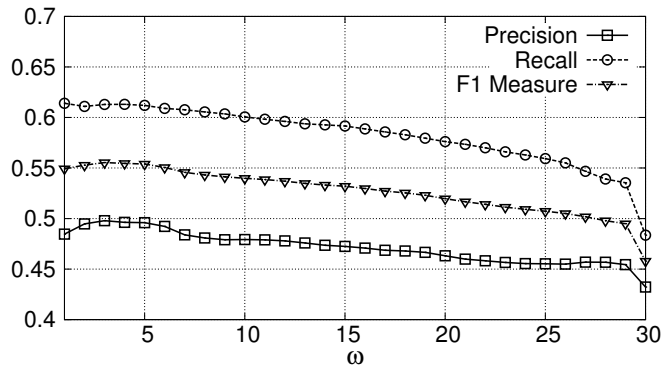}\medskip
\label{fig:face-omega3}
}\caption[]{
Accuracy of the three types of detectors when the detection results are accumulated over $\omega$ frames.
The ground truth was labeled manually for a one minute video.
Precision, recall and F1 measure are averaged over all pixels of all frames. The optimal window size if found to be $\omega = 4$.
}
\label{fig:omega}
\end{figure*}

We analyzed the accuracy of the three detectors for different window sizes $\omega$.
The optimal value is a trade-off between minimizing the influence of spurious false detections and motion blur in the detection result.
Fig. \ref{fig:omega} shows the accuracy of the three detection methods for varying values of $\omega$.
The accuracy is expressed in terms of precision, recall, and F1 measure.
The ground truth was obtained by manually tagging the regions in a one minute test video.
The accuracy is measured on the pixel level.
For each pixel, we compare the binary detection result with the binary ground truth.
Precision, recall and F1 measure are calculated by counting correct detections, false positives and false negatives over the entire size of the frame and averaging them over the one minute video.
From the obtained data for all three detectors, we derived $\omega=4$ as a suitable window size. The optimal window size does not have to be the same for all detectors. For a different set of semantic observations or a different surveillance site, the optimal window size can be different.

\subsection{Timing Analysis} \label{Timing Analysis}
\begin{figure*}
\centering
\subfigure[Face]
{
\includegraphics[width = 0.5\linewidth]{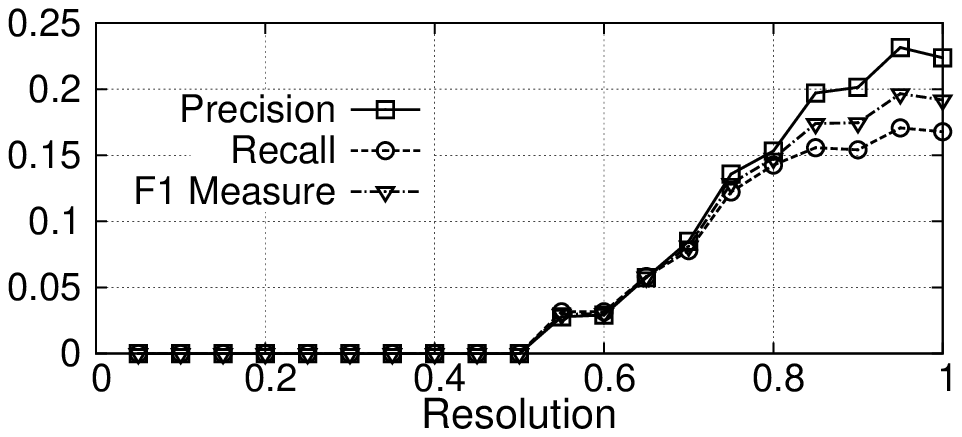}\medskip 
\label{fig:face-resolution1}
}
\subfigure[Human]
{
\includegraphics[width = 0.5\linewidth]{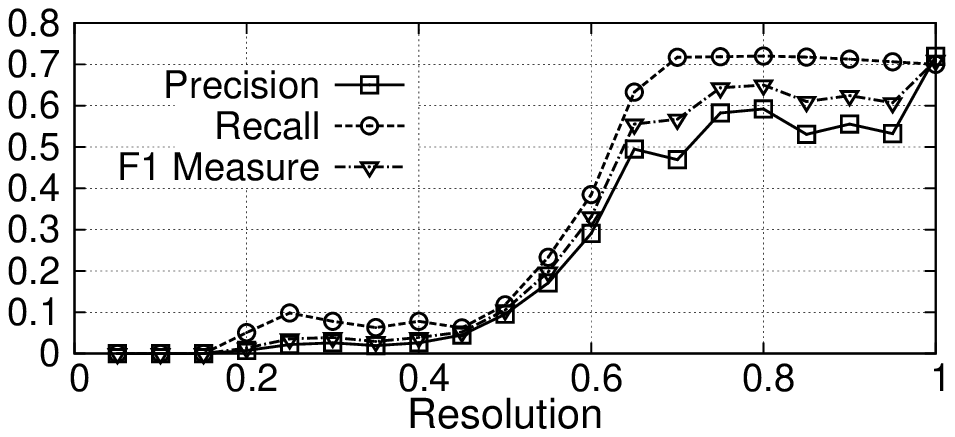}\medskip
\label{fig:face-resolution2}
}
\subfigure[Motion]
{
\includegraphics[width = 0.5\linewidth]{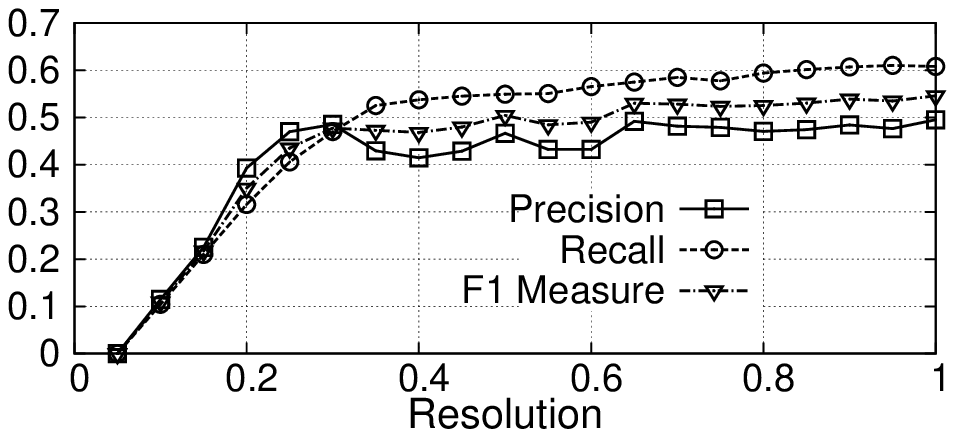}\medskip
\label{fig:face-resolution3}
}\caption[]{
Accuracy of the three types of detectors for different scale factors of the input video.
The face detection accuracy decreases linearly with resolution, while the accuracy of body and foreground detection remains nearly constant for scale factors down to 0.8 and 0.6, respectively.}
\label{fig:resolution}
\end{figure*} 

In the timing experiment, we vary the resolution of the frame that is used as input to the detection methods and measure the resulting detector accuracy.
The goal is to save processing time by reducing the resolution as much as possible while maintaining the accuracy.
The ground truth and the measurement of accuracy is identical to the previous experiment.
Figure \ref{fig:resolution} shows the results of the experiment. 

The face detection accuracy decreases linearly with scale, as shown in Figure \ref{fig:face-resolution1}.
We thus do not reduce the image resolution for the face detector.
In the case of human and motion detection, however, the accuracy remains nearly constant initially and then decreases rapidly when the images are too small.
For the human detector, there is no significant decrease in accuracy until the frame size is scaled down from full HD to a factor of 0.8, as can be seen in Figure \ref{fig:face-resolution2}.
Similarly, the resolution of the input frame of the motion detector can be reduced by a factor of 0.6 (Figure \ref{fig:face-resolution3}).
For both detectors, the decrease in F1 value is less than $0.05$.

\subsection{Accuracy and Efficiency}

\begin{figure*}
\centering
\subfigure[Accuracy]
{
\includegraphics[width = 0.6\linewidth]{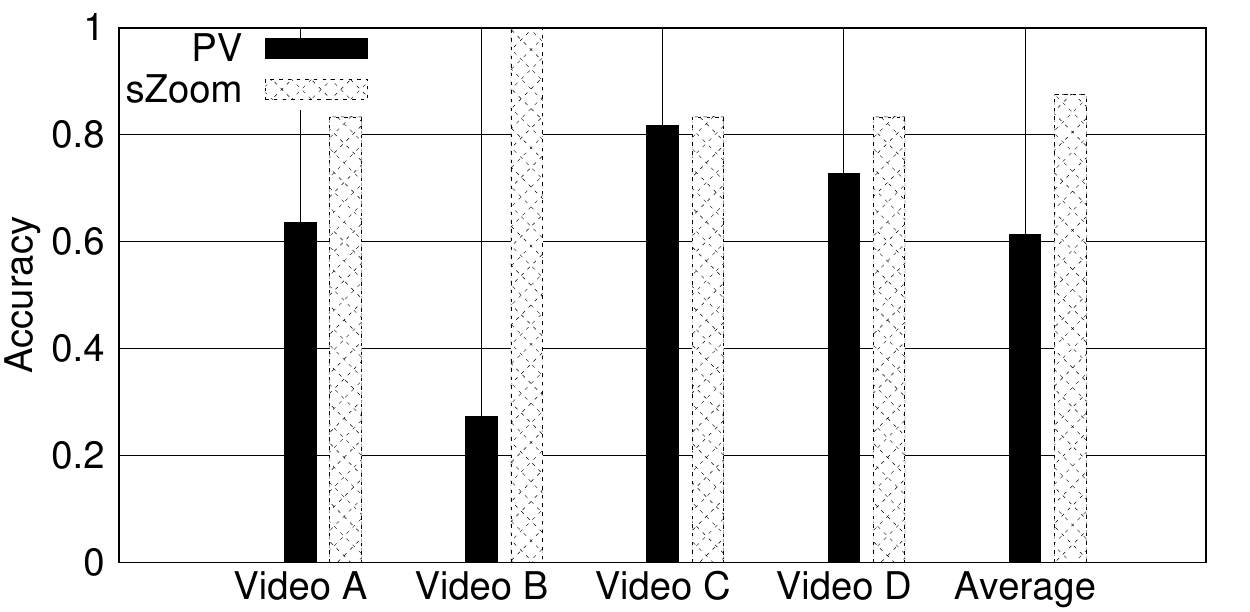}\medskip
\label{fig:accuracy-imp}
}
\subfigure[Frames per second]
{
\includegraphics[width = 0.6\linewidth]{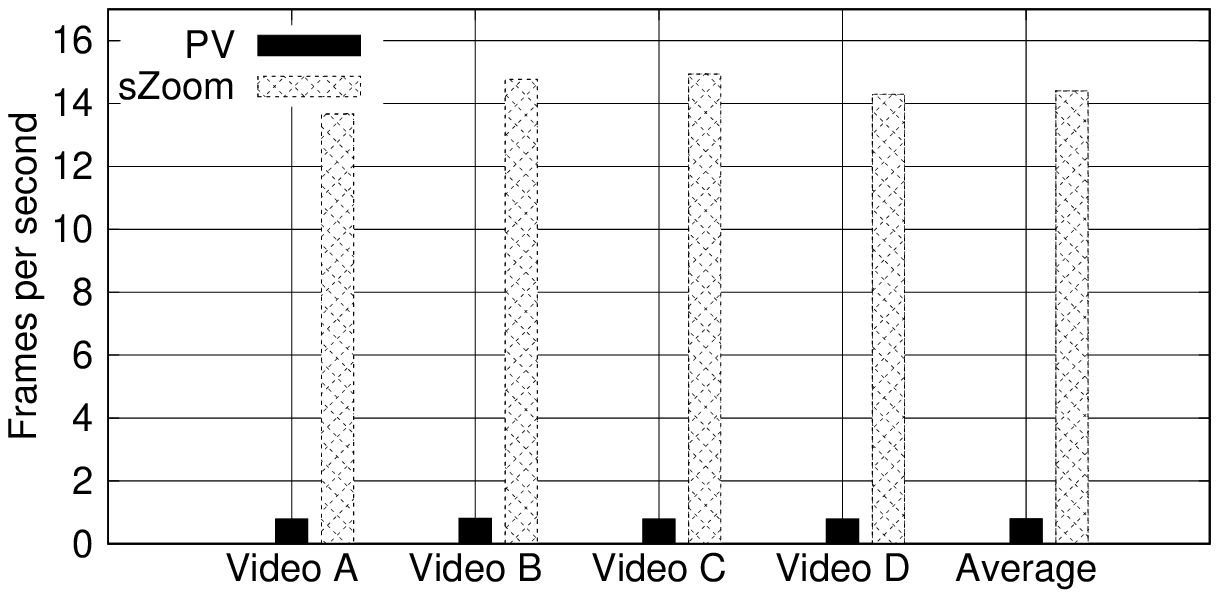}\medskip
\label{fig:time-imp}
}\caption[]{
Comparison between the proposed framework (sZoom) and the preliminary version (PV) of the system for four 1 minute videos.
(a) compares the accuracy, which is the number of correct zoom operations divided by the total number of zooms.
In (b), the average frames per second are compared.
The large increase in the frames per second is due to the reduced accumulation window size $\omega$ and the reduced image resolution.
}
\label{fig:imp}
\end{figure*} 

With the incorporation of tracking and refinements in the framework, we are able to improve the performance of sZoom compared to the preliminary version of the work \cite{kuang2014real}. The improvement is two-fold: reduced processing time and improved accuracy. We have already discussed the improvement in latency. The other improvements are reported in terms of processing time per frame and  accuracy.
In this experiment, the accuracy is measured as the ratio between the number of times the system zooms into the correct ROI and the total number of zoom operations (i.e., cycles). The correct ROI corresponds to the objects and/or persons inside the ROI chosen at the beginning of the cycle. If the object and/or persons are still there in the chosen output region by the end of the stay period, the system has zoomed into the correct ROI, otherwise not. 
The results are shown in Figure \ref{fig:accuracy-imp}.
The accuracy improved by 42\% over the preliminary version. 
The processing time of the proposed system is also reduced in comparison to the preliminary version.
Due to the reduced accumulation window size $\omega$, the costly human detection only needs to be performed on 2\% of the frames. In addition, human and motion observations are made on the lower resolution images. With these optimizations, processing time was reduced to 5\% of the original time taken by the baseline system. The improvement is shown in Fig. \ref{fig:time-imp} in terms of frames per second. 
The frames per second is measured  by dividing the  total number of frames by time taken to process the entire video. Four videos of 1 minute each are used in the experiment. 

\subsection{Qualitative User Study} \label{Qualitative User Study}
\begin{figure*}
\centering
\subfigure
{
\includegraphics[width = \widthone \linewidth]{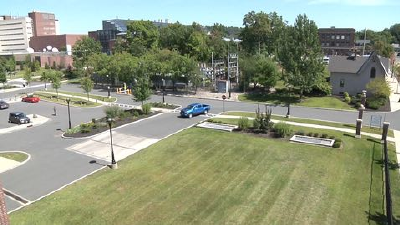}\medskip \hspace{-0.2cm}
}
\subfigure
{
\includegraphics[width = \widthone\linewidth]{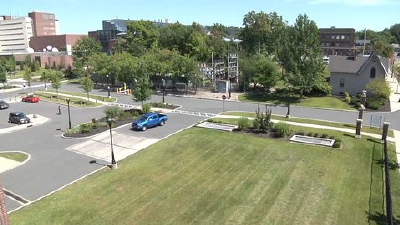}\medskip \hspace{-0.2cm} 
}
\subfigure
{
\includegraphics[width = \widthone\linewidth]{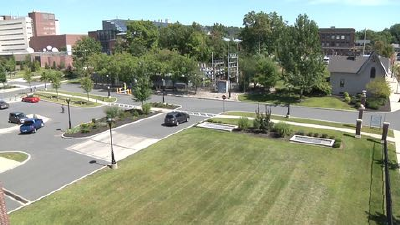}\medskip \hspace{-0.2cm}
}
\subfigure
{
\includegraphics[width = \widthone\linewidth]{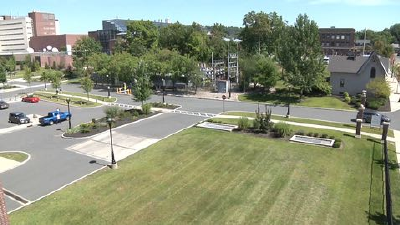}\medskip \hspace{-0.2cm}
}
\subfigure
{
\includegraphics[width = \widthone\linewidth]{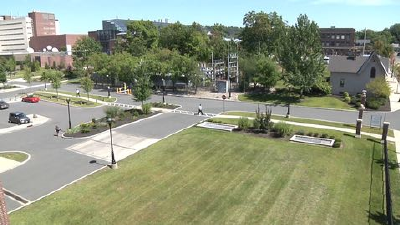}\medskip \hspace{-0.2cm}
}
\subfigure
{
\includegraphics[width = \widthone\linewidth]{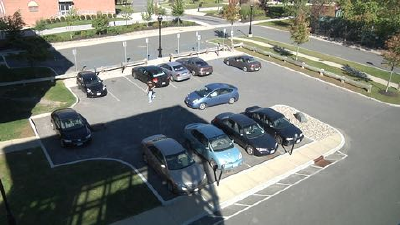}\medskip \hspace{-0.2cm}
}
\subfigure
{
\includegraphics[width = \widthone\linewidth]{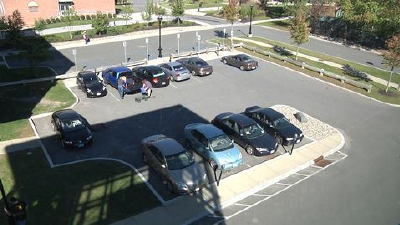}\medskip \hspace{-0.2cm}
}
\subfigure
{
\includegraphics[width = \widthone\linewidth]{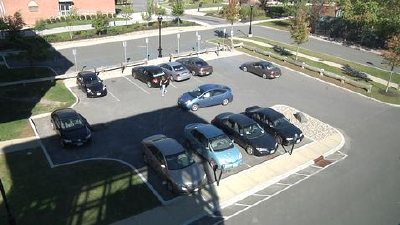}\medskip \hspace{-0.2cm}
}
\subfigure
{
\includegraphics[width = \widthone\linewidth]{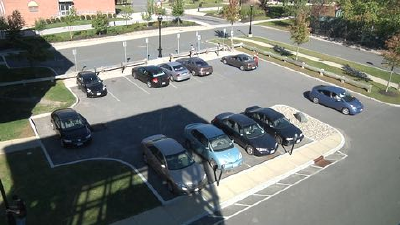}\medskip \hspace{-0.2cm}
}
\subfigure
{
\includegraphics[width = \widthone\linewidth]{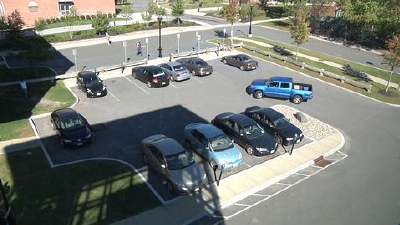}\medskip \hspace{-0.2cm}
}

\caption[]{
Representative frames from video A (top row) and video B (bottom row) used in the user studies.
The videos are taken from the VIRAT activity recognition dataset~\cite{oh2011large}.
}
\label{fig:videoA}
\end{figure*} 

%
%

\begin{figure*}
\centering
\subfigure[]
{
\includegraphics[width = \widthtwo\linewidth]{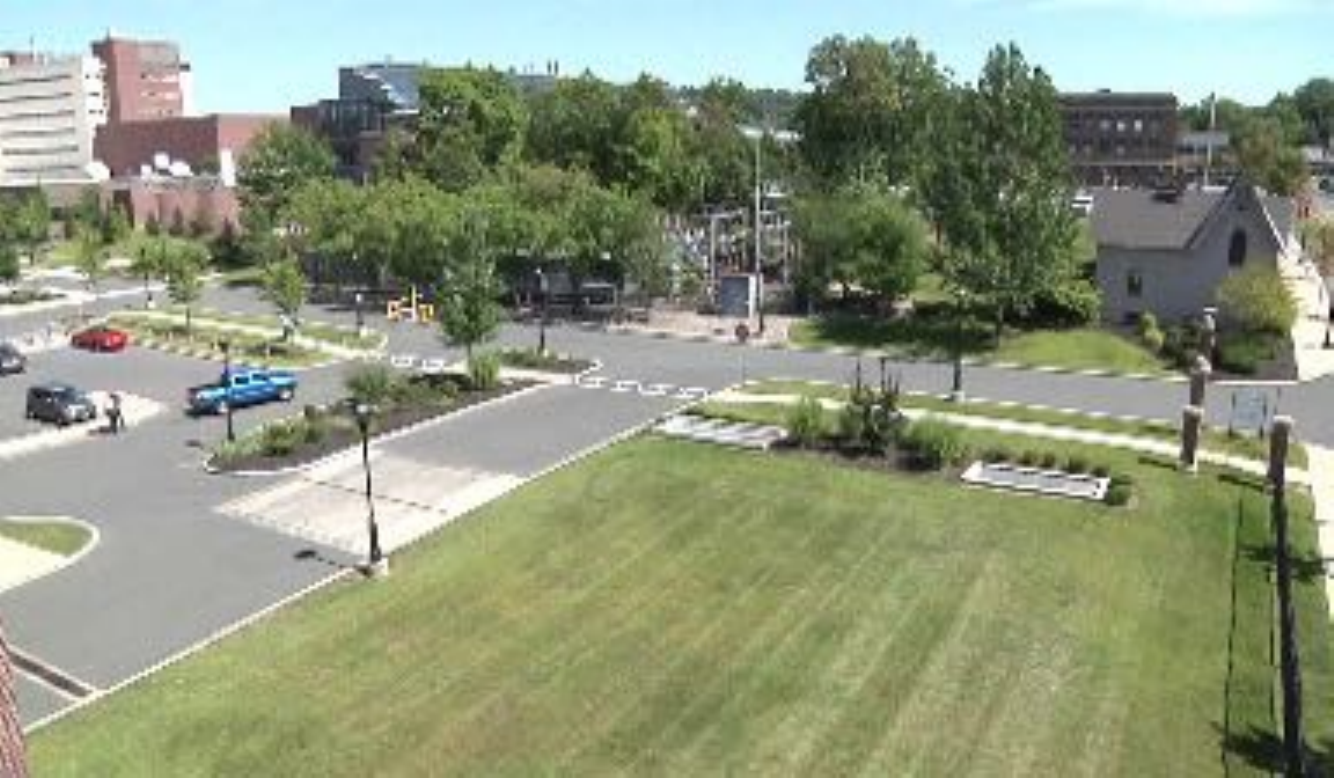}\medskip \hspace{-0.2cm} 
}
\subfigure[]
{
\includegraphics[width = \widthtwo\linewidth]{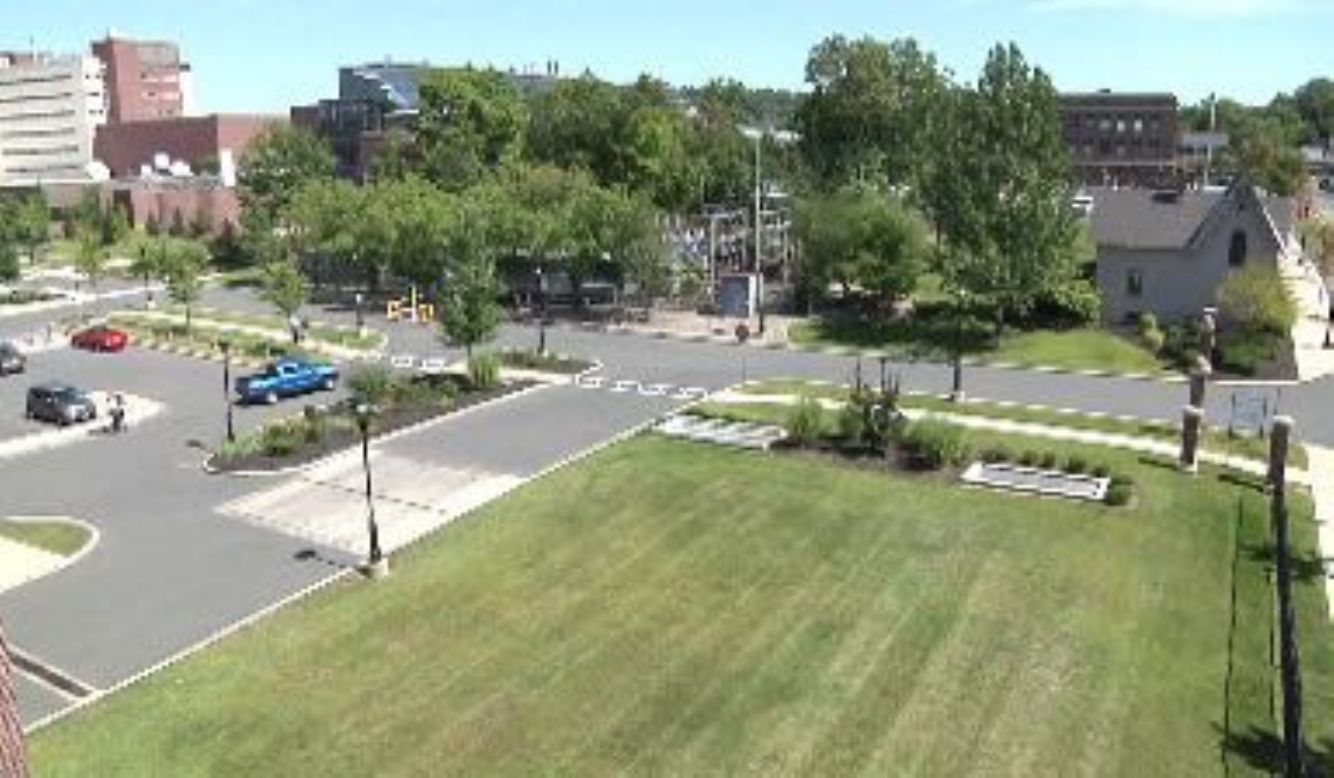}\medskip \hspace{-0.2cm}
}
\subfigure[]
{
\includegraphics[width = \widthtwo\linewidth]{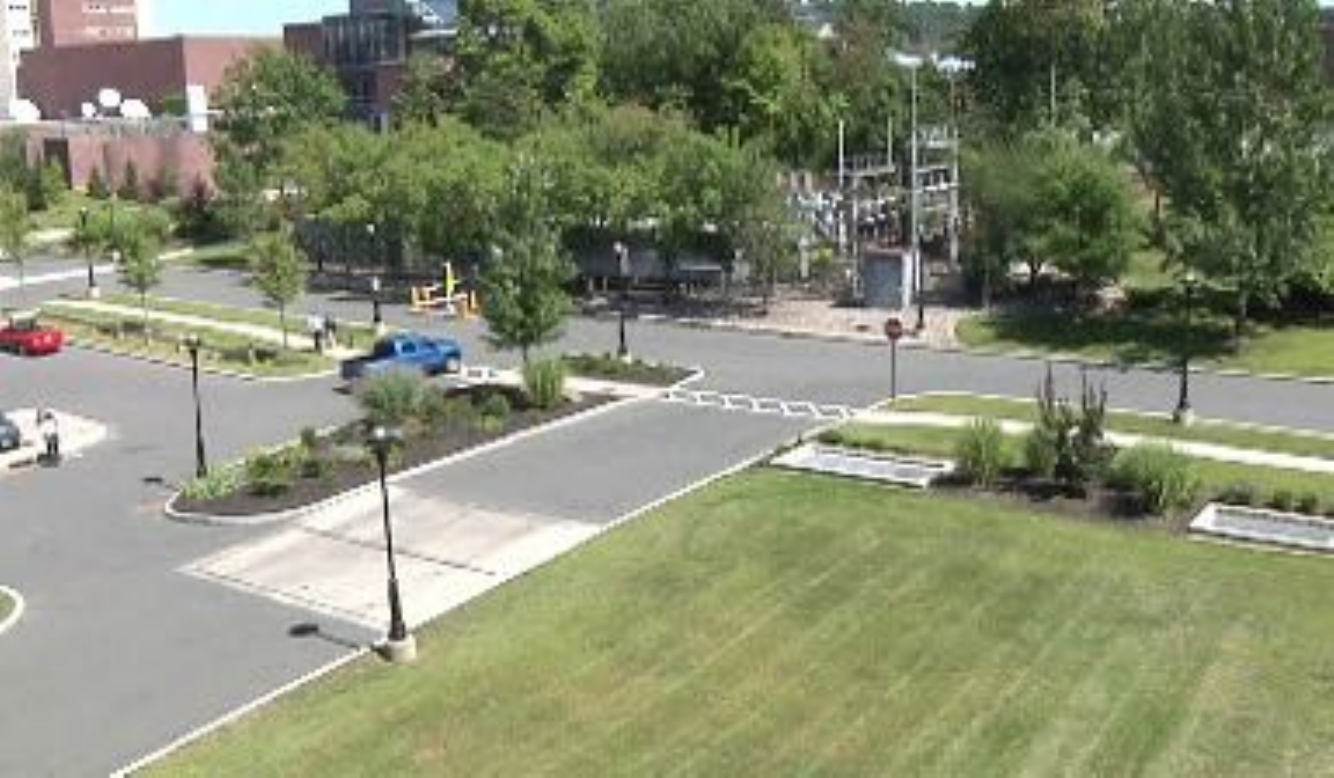}\medskip \hspace{-0.2cm}
}
\subfigure[]
{
\includegraphics[width = \widthtwo\linewidth]{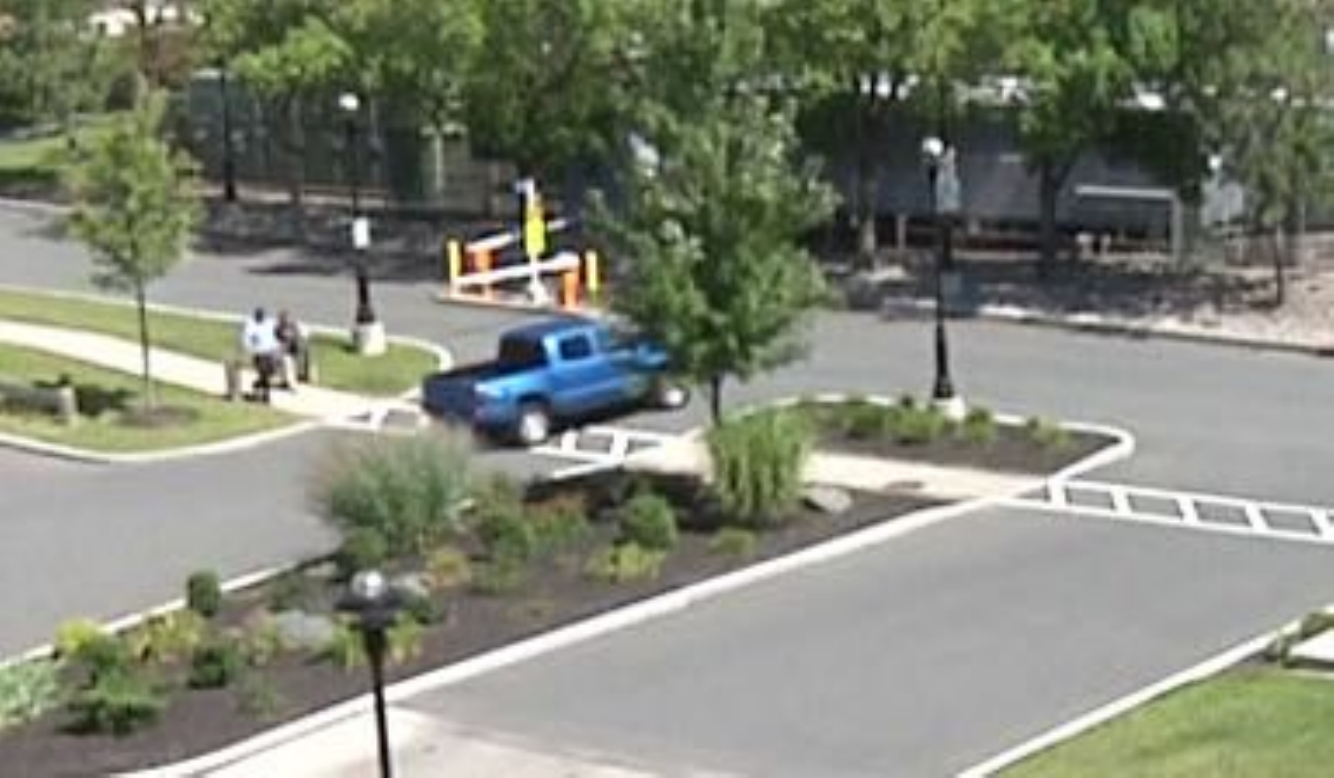}\medskip \hspace{-0.2cm}
}
\subfigure[]
{
\includegraphics[width = \widthtwo\linewidth]{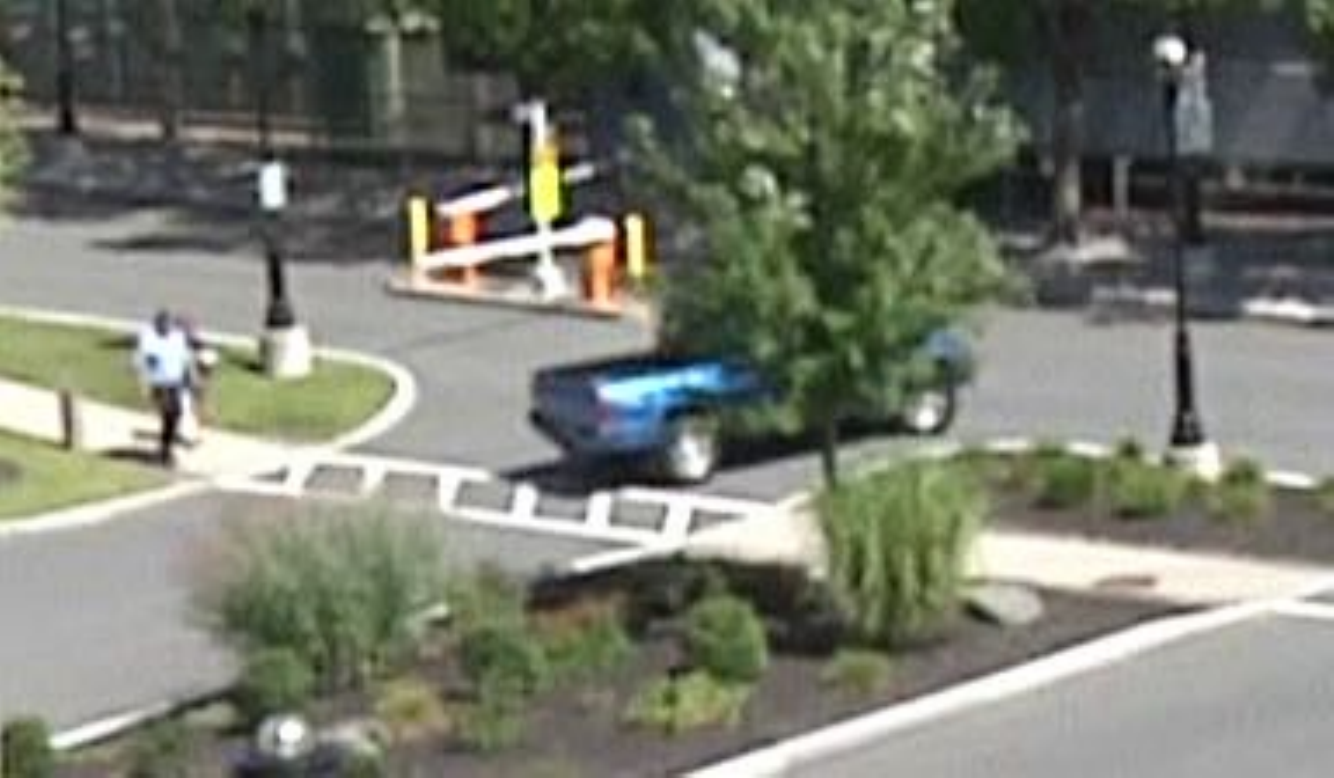}\medskip \hspace{-0.2cm}
}
\subfigure[]
{
\includegraphics[width = \widthtwo\linewidth]{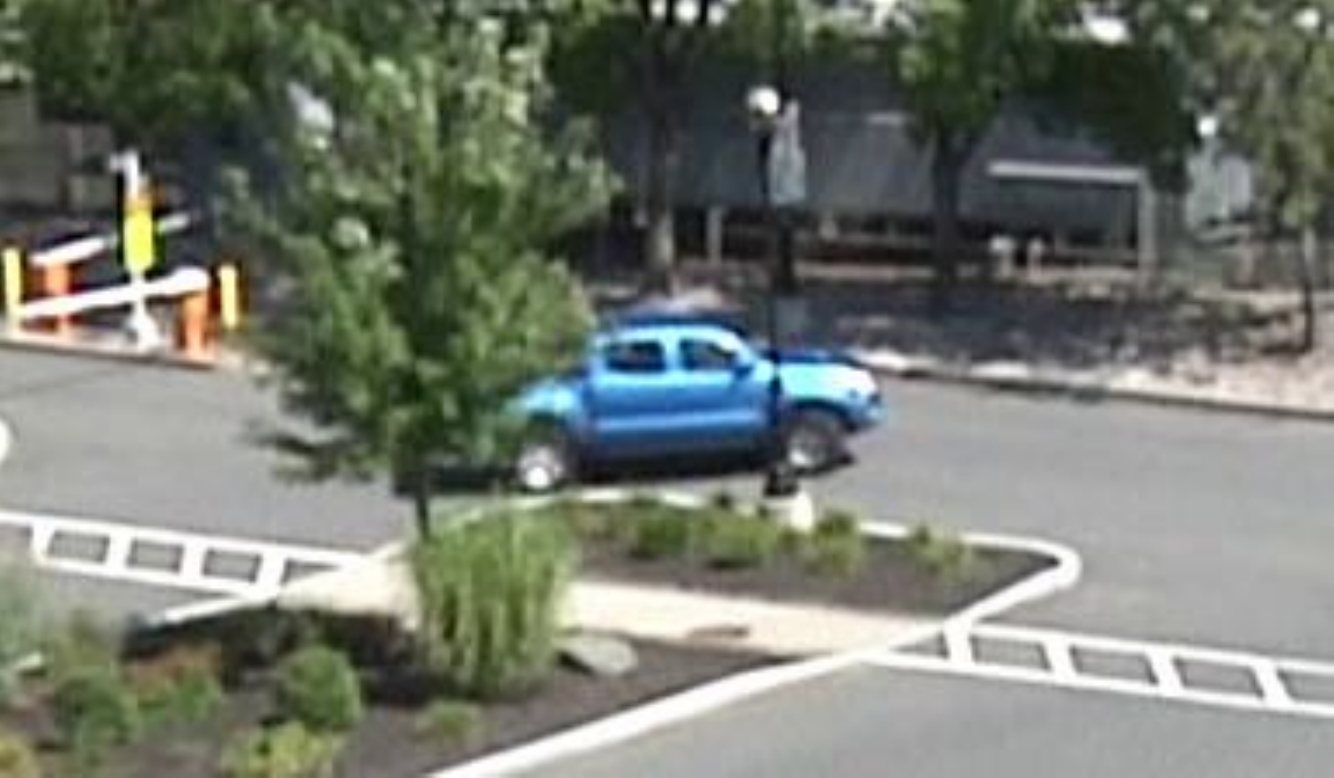}\medskip \hspace{-0.2cm}
}

\subfigure[]
{
\includegraphics[width = \widthtwo\linewidth]{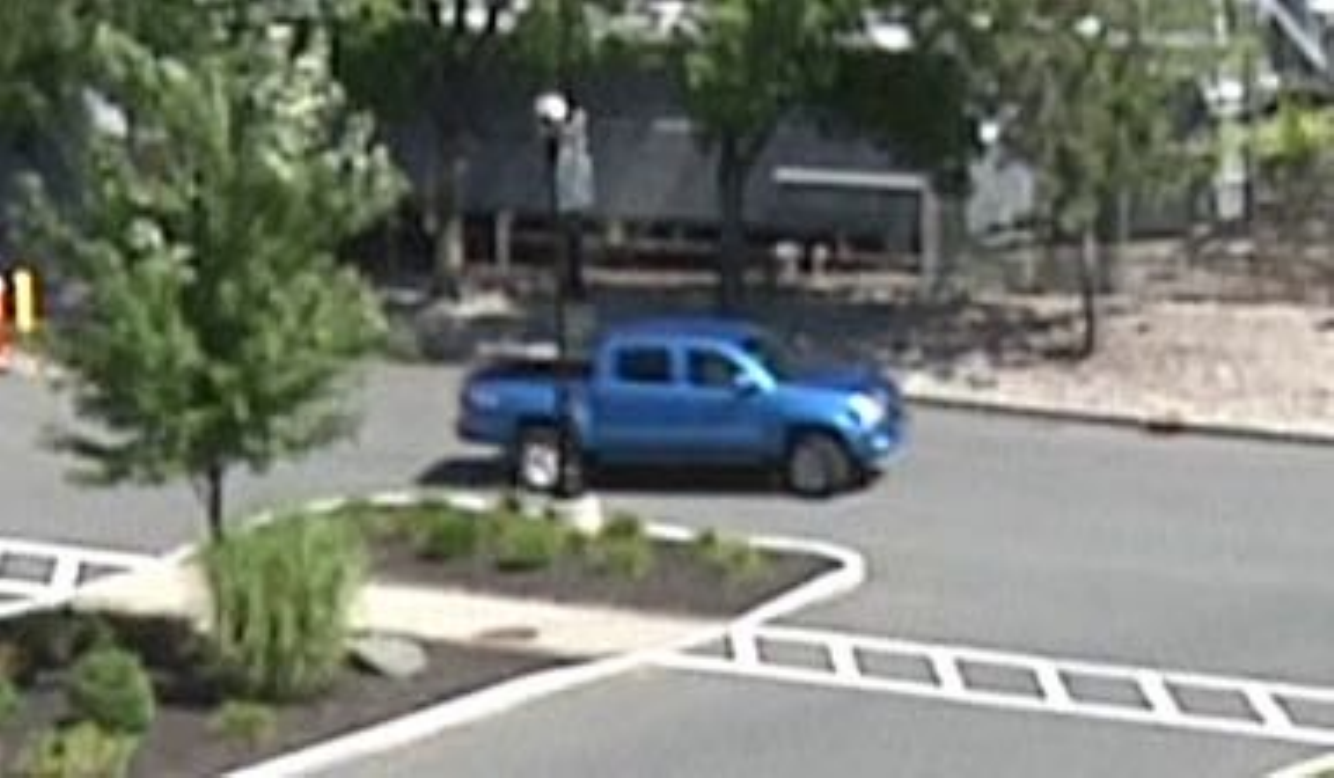}\medskip \hspace{-0.2cm}
}
\subfigure[]
{
\includegraphics[width = \widthtwo\linewidth]{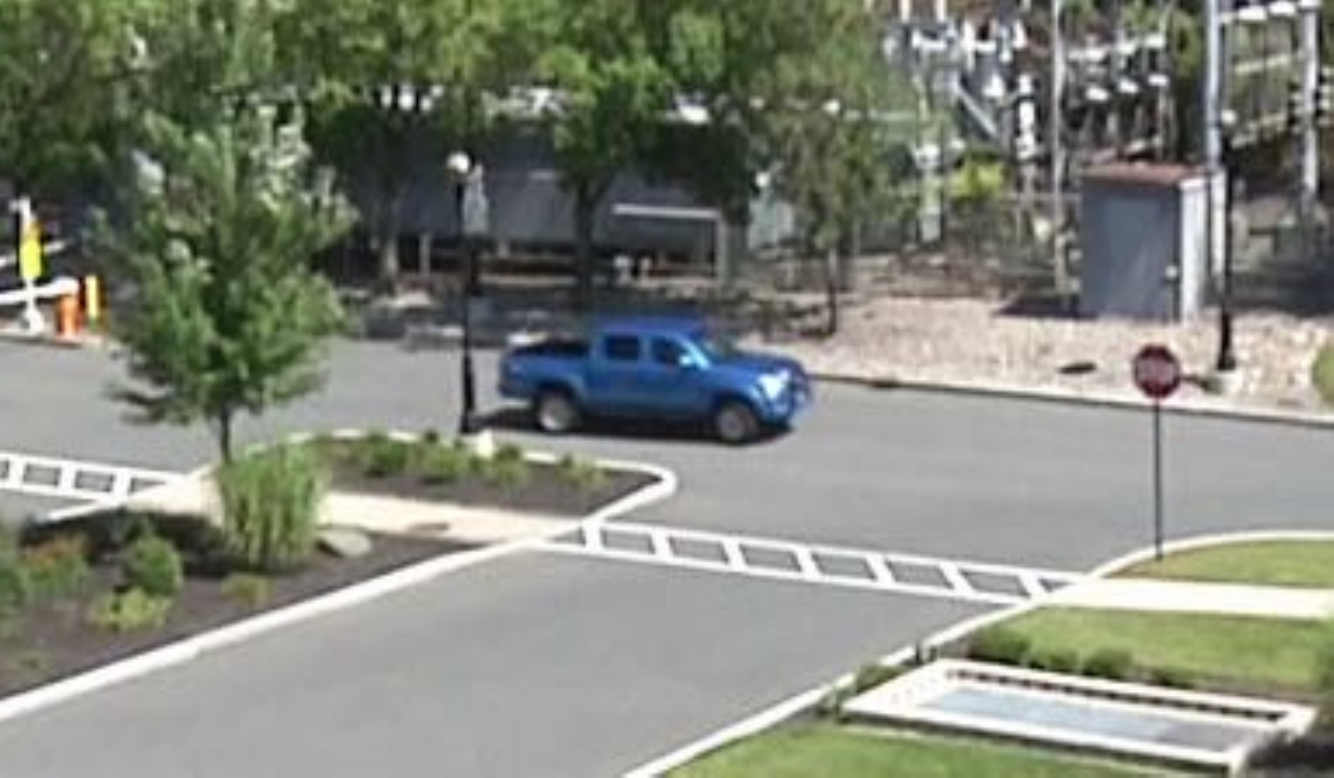}\medskip \hspace{-0.2cm}
}
\subfigure[]
{
\includegraphics[width = \widthtwo\linewidth]{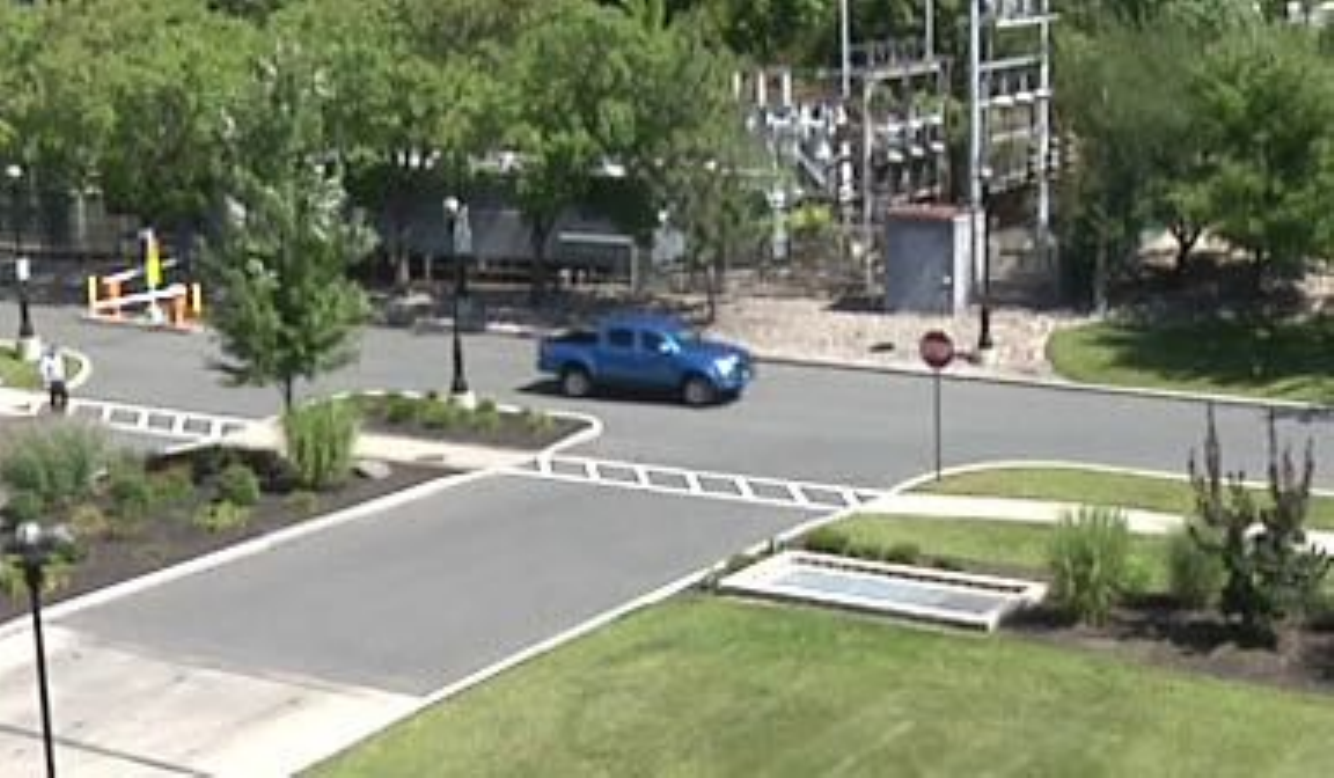}\medskip \hspace{-0.2cm}
}
\subfigure[]
{
\includegraphics[width = \widthtwo\linewidth]{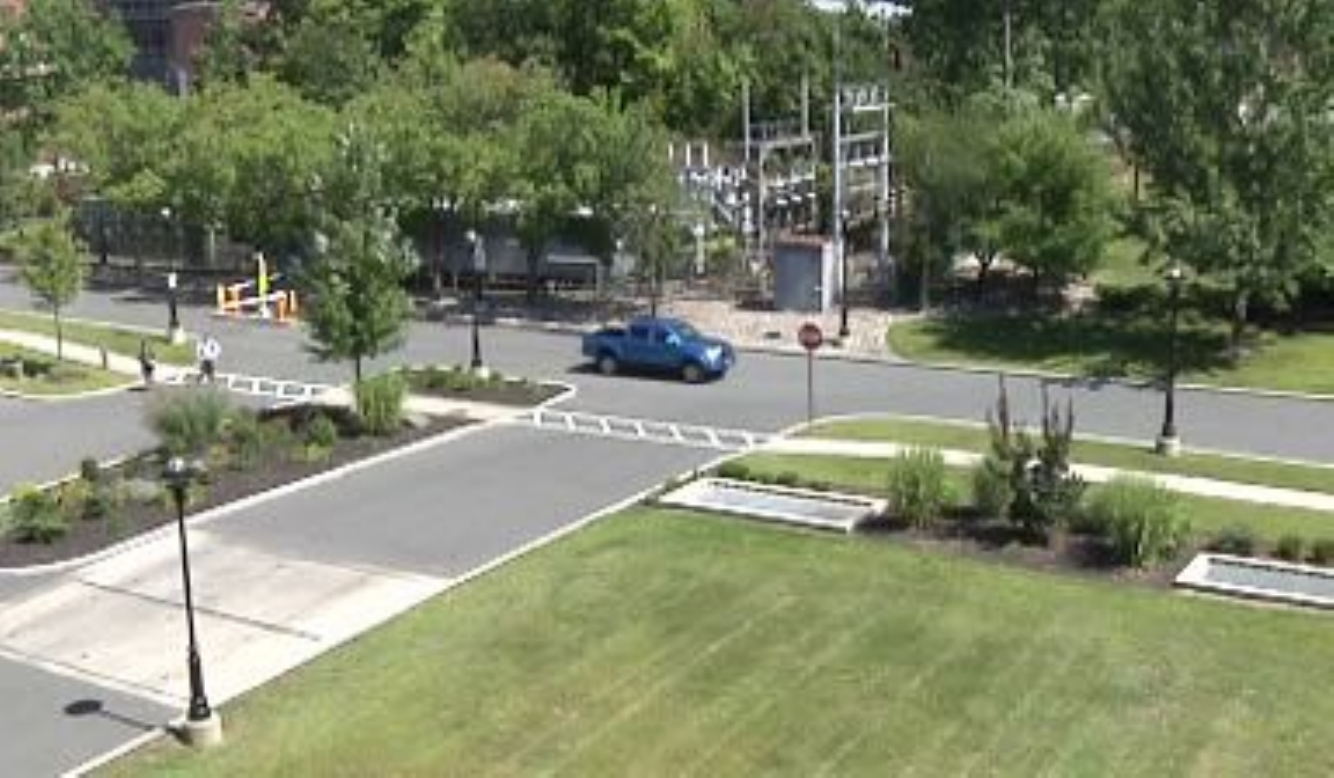}\medskip \hspace{-0.2cm}
}
\subfigure[]
{
\includegraphics[width = \widthtwo\linewidth]{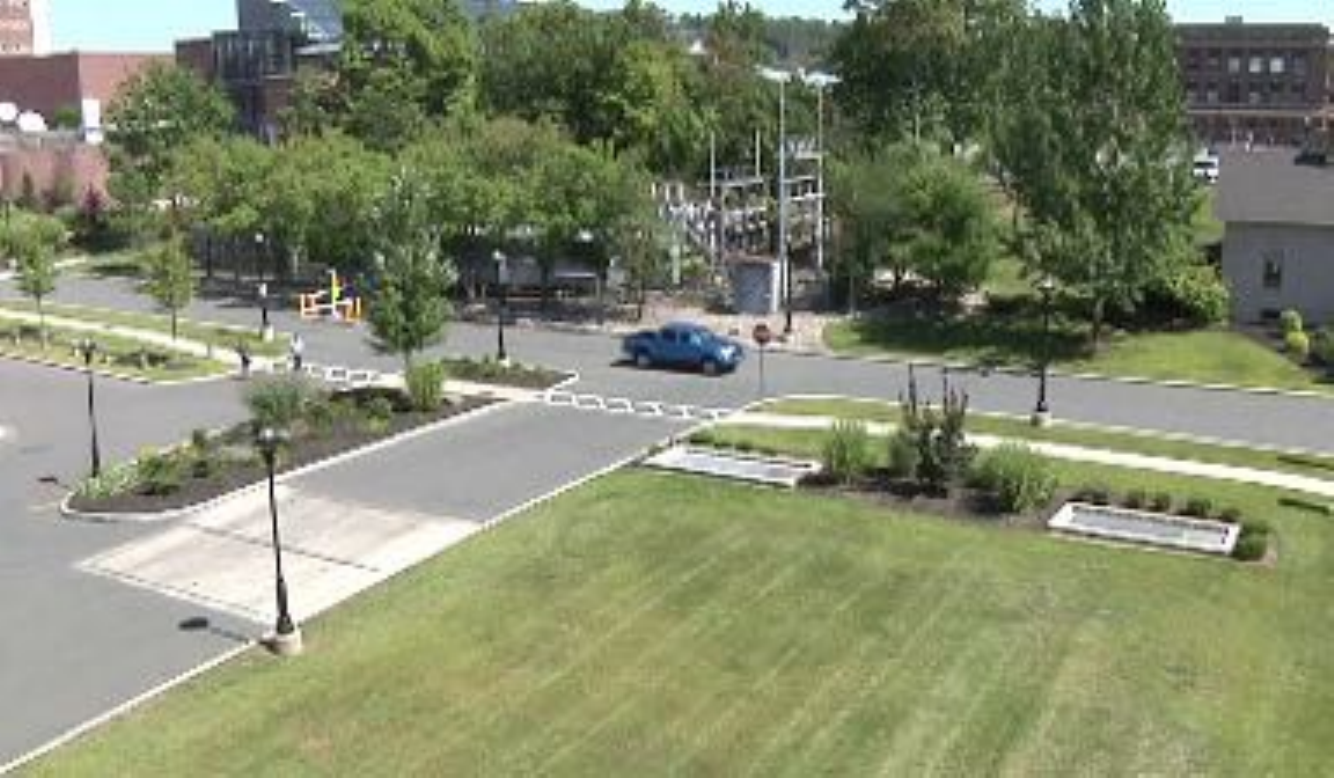}\medskip \hspace{-0.2cm}
}
\subfigure[]
{
\includegraphics[width = \widthtwo\linewidth]{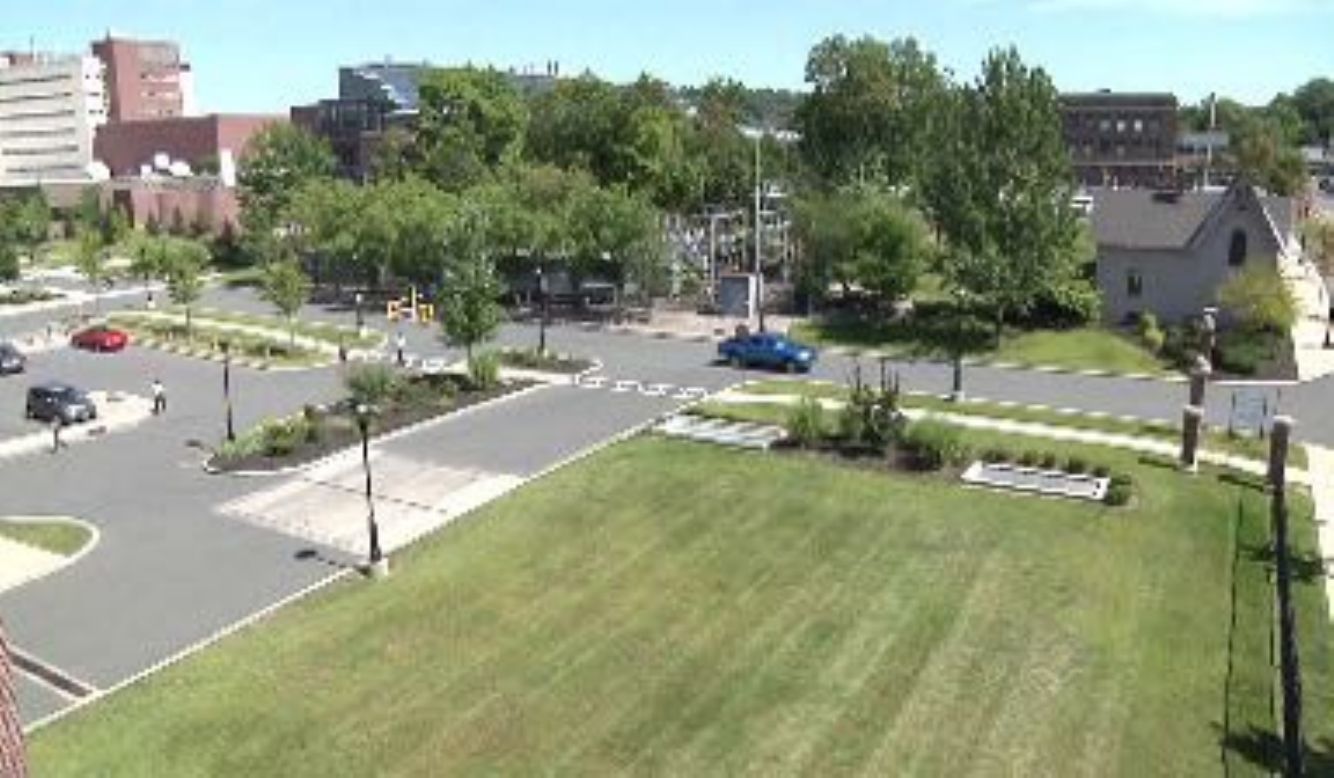}\medskip \hspace{-0.2cm}
}

\caption[]{
Exemplary output of sZoom.
Fig. a, b show zoomed-out stage. The system detects the car as a salient object and starts zooming-in (Fig. c, d, e).
As the car is moving, the system tracks and zooms to keep the car in the center of the zoomed view  (Fig. f, g, h).
The system zooms-out again to provide the user with contextual information (Fig. i, j, k). Fig. l again show the zoomed-out stage. 
}
\label{fig:videoA-Output}
\end{figure*}

The goal of the qualitative user study is to show that the videos produced using sZoom are more appropriate for surveillance than the original videos when viewed on a small screen.  
We use two HD videos ($1920 \times 1080$, 30 fps, 60 seconds) from the VIRAT dataset \cite{oh2011large} which is designed for the assessment of activity recognition algorithms.
We refer to them as video A and video B.
Representative frames from these videos are shown in Figure~\ref{fig:videoA}.
The target resolution is chosen as $384 \times 216$, which is one fifth of the input resolution. 

The parameter $\omega$ was set to the value of 4 that was determined experimentally in Section~\ref{Selection of Accumulation Window Size}.
The cycle length $\Delta=5 s$ was chosen empirically for user's comfort.
A value that is too low results in a video that appears rushed whereas a value that is too large may lead to the missing of important events. The decay factor $\alpha$ for the multivariate Gaussian map is chosen as $\alpha=0.3$ according to the amount of activity in the scene.
Its value should be low for high activity sites and vice-versa.
The optimal conformal coefficients of the three detectors were calculated as $c_m=0.46$, $c_h=0.53$, and $c_f=0.01$ according to their detection reliability for the given scenes.
An example output frame sequence for video A is shown in Figure~\ref{fig:videoA-Output}.
The systems zooms into a moving car, tracks the car and then zooms out again.   

The user study is conducted online through a website\footnote{\url{https://sites.google.com/site/userstudytype1/}}.
It shows the output video produced with sZoom and the original video scaled down to the target resolution side-by-side.
In the remainder of the experiments section, ``video A-D'' refers to the different input videos, i.e., the different scenes under surveillance.
The term ``version'' refers to the approach that was used to produce the output video, i.e., scaling or sZooming.
To avoid bias, we randomize the position (left or right) at which the versions are presented for each user.
Users are asked to take on the role of a security operator whose main task is to monitor a surveillance site and observe important activities in order to detect any abnormal situations.
Five statements were given and the users had to rate their agreement on a scale from 1 to 5.
See the leftmost column of Table \ref{table:results-qualitative} for a list of the statements.
The statements are designed to assess whether or not the video provides a detailed view of the site (statement 1 and 2), reduces operator boredom (3), and is helpful in monitoring activities (4 and 5).


A total of 54 users (42 males, 12 females) rated video A, and 45 users rated video B (32 males, 13 females).
Most of the users aged between 25 and 35. The resulting average ratings for all five statements for the two versions of each of the two videos are shown in Table \ref{table:results-qualitative}.
The value in parentheses is the standard deviation.
The low standard deviation indicates agreement among the viewers.
Users consistently gave higher ratings to the proposed version that it provides details of the site (1 and 2), reduces the operator boredom (3), and is helpful in monitoring activities (4 and 5).
It can be seen from the results that the users find the proposed output videos more appropriate for surveillance than a scaled version of the original video.
Across both videos and all questions, the users gave 46\% higher ratings to the sZoomed version than to the scaled one. 

In order to ensure that the rating are not higher just by chance, we conduct one-tailed t-test for paired samples. We choose one-tailed t-test because we expect the sZoomed rating to always be higher than the scaled video. The statistics are calculated across videos for each question. The threshold for statistical significance, i.e. alpha, is chosen as 0.05. The null hypothesis is that the true difference in mean ratings is zero. Since there are 99 user responses in total, the degree of freedom is 98. The corresponding $t_{stat}$, $t_{cric}$ are given in the last three columns of Table \ref{table:results-qualitative}. We can see that the observed t statistics is significantly higher than the critical value corresponding to alpha = 0.05. Hence, the probability of observing the given sample with equal means is almost zero and the null hypothesis is rejected. Hence, the means are indeed different, and sZoom ratings are higher.


\begin{table*}
\centering
\caption{Average user ratings for five statements and two videos.
The proposed method is compared to a video that was scaled down to the target size. The values in parentheses are standard deviation.}
\begin{tabular}{|m{5cm}|m{1.3cm}|m{1.3cm}|m{1.3cm}|m{1.3cm}|m{1cm}|m{1cm}|}\hline
\multicolumn{1}{|c|}{\multirow{2}{*}{Statement}}&\multicolumn{2}{c|}{Video A}& \multicolumn{2}{c|}{Video B} &\multicolumn{2}{c|}{t-Test}
\\\cline{2-7}
& Scaled  & \textbf{sZoomed}  &Scaled  & \textbf{sZoomed}& $t_{stat}$ & $t_{cric}$ \\
\hline
1: The video is able to provide details of the regions that are important for surveillance. &	2.8 (1.1)  &	\textbf{4.3 (1.0)} &	2.9 (1.0)	 &\textbf{4.5 (0.7)} & 10.5 & 1.7\\
\hline
2: The video provides a detailed view of all important events and activities happening at the site. &	3.0 (1.1)	 &\textbf{4.2 (0.9)}	  &2.8 (1.0)  &	\textbf{4.3 (0.8)} & 8.6 & 1.7 \\
\hline
3: The presentation of the video reduces boredom and keeps me engaged in the activities in the video. &	2.8 (1.0) &	\textbf{4.1 (1.0)} &	2.7 (1.0) &	\textbf{4.3 (0.7)} & 9.7 & 1.7 \\
\hline
4: The presentation of the video is helpful in monitoring the area. &	3.2 (1.0) &\textbf{4.2 (1.0)} &	3.0 (0.9) &	\textbf{4.2 (0.8)} & 8 & 1.7\\
\hline
5: It is easy to understand the activities in the video.	 &2.9 (0.8) &	\textbf{4.2 (0.9)} &	3.1 (1.0) &	\textbf{4.2 (0.8)} & 9.97 & 1.7\\
\hline
\multicolumn{1}{|c|}{\textbf{Average}} & 2.9 (1.02) & \textbf{4.2 (0.97)} & 2.9 (1.01) & \textbf{4.3} (0.77)\\
\hline
\end{tabular}
\label{table:results-qualitative}
\end{table*}

\subsection{Quantitative User Study}

\begin{figure*}
\centering
\subfigure
{
\includegraphics[width = \widthone\linewidth]{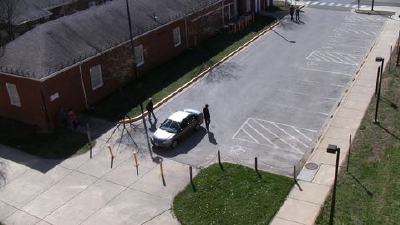}\medskip \hspace{-0.2cm}
}
\subfigure
{
\includegraphics[width = \widthone\linewidth]{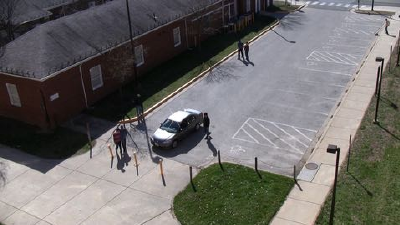}\medskip \hspace{-0.2cm} 
}
\subfigure
{
\includegraphics[width = \widthone\linewidth]{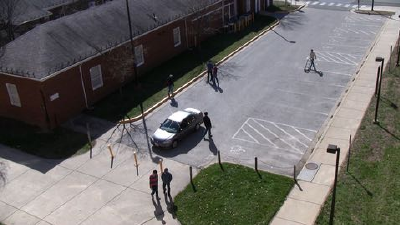}\medskip \hspace{-0.2cm}
}
\subfigure
{
\includegraphics[width = \widthone\linewidth]{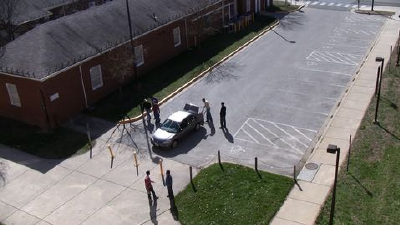}\medskip \hspace{-0.2cm}
}
\subfigure
{
\includegraphics[width = \widthone\linewidth]{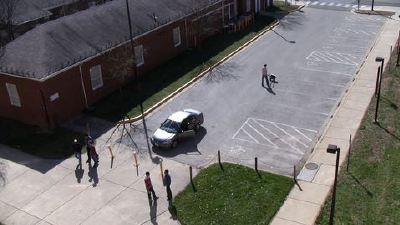}\medskip \hspace{-0.2cm}
}
\subfigure
{
\includegraphics[width = \widthone\linewidth]{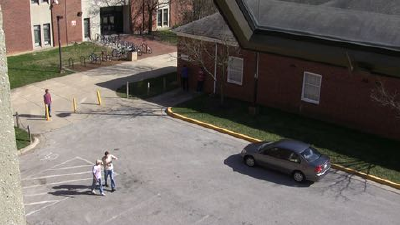}\medskip \hspace{-0.2cm}
}
\subfigure
{
\includegraphics[width = \widthone\linewidth]{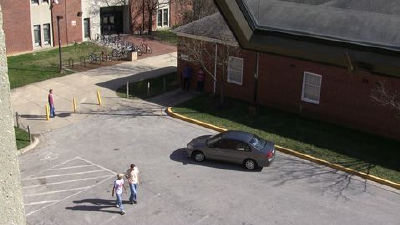}\medskip \hspace{-0.2cm}
}
\subfigure
{
\includegraphics[width = \widthone\linewidth]{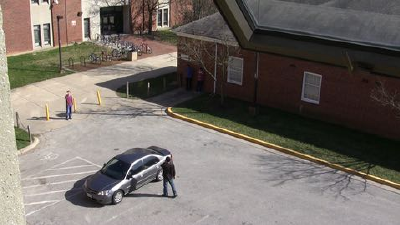}\medskip \hspace{-0.2cm}
}
\subfigure
{
\includegraphics[width = \widthone\linewidth]{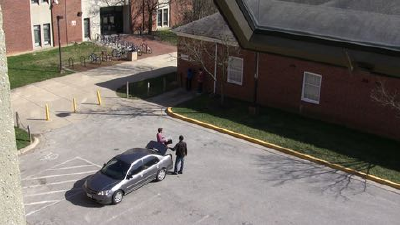}\medskip \hspace{-0.2cm}
}
\subfigure
{
\includegraphics[width = \widthone\linewidth]{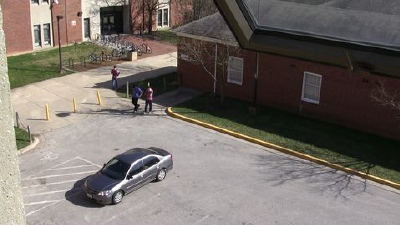}\medskip \hspace{-0.2cm}
}

\caption[]{
Representative frames from video C (top row) and video D (bottom row) used in the user studies.
The videos are taken from the VIRAT activity recognition dataset~\cite{oh2011large}.
}
\label{fig:videoCD}
\end{figure*}

In the previous experiment, the users agreed that sZoom provides a more detailed view of the scene and reduces operator boredom.
To quantitatively measure how much information the users gain from the two versions of the videos, we performed a second study.
We asked the users to answer questions about the activities in the scene to test how many of the events they noticed and remembered.
Two separate videos named C and D taken from the VIRAT dataset~\cite{oh2011large} are used this time. Both videos have HD resolution at 30 frames per second. Representative frames from these two videos are shown in Figure  \ref{fig:videoCD}. We used the same parameter values that are used  in the previous experiment to produce the output videos. 

In order to avoid undesired bias in the user responses, the following measures were taken:
\begin{itemize}
\item If one user watches both the scaled and the proposed version of the same video, they would find it easier to remember details of the version they watch later. Thus, one user only gets to watch one version of a given video.  
\item Approximately half the users watched the scaled version of C while the other half watched the proposed version. They then watched the version of D that was produced by the \emph{other} approach.
\item The users were only allowed to answer the questions after having watched the video entirely. They were not allowed to watch the videos more than once.
\item Because the video thumbnail itself reveals certain details, the questionnaire was put on a separate page which the users could access after watching the video.
\end{itemize}   
As a consequence of the first two items, approximately half the users watched ``scaled C and proposed D'' while the other half watched ``proposed C and scaled D''.
The questions were specific to the content of the input videos.
Therefore, there are two sets of questions: three questions for video C and six questions for video D.
All user responses are checked manually for correctness.
The number of users who viewed the scaled version of a video was different from the number of users who viewed the sZoomed version. In order to estimate and compare the amount of information that was gained by watching a version of a video, we define the term \emph{Information Quotient} ($IQ$) for each question and each version as follows:

\begin{equation}
IQ = \frac{\mbox{Number of correct responses}}{\mbox{Total number of responses}}
\end{equation} 

The number of users who watched ``scaled C and proposed D'' was 18 (7 female and 11 male).
``proposed C and scaled D'' was watched by 26 (6 female and 20 male) users.
All the users aged between 20 and 35.
The resulting $IQ$ values for each question for video C and video D are given in Table \ref{table:results-quantativeC} and Table \ref{table:results-quantativeD} respectively.
We can see that the $IQ$ across all questions is higher for the proposed version than the scaled one.
This means that users are able to extract and retain more details of the scene by watching the output video produced by sZoom.

The difference in $IQ$ is largest for the questions related to small details.
For example, in the beginning of video C, the person by the car makes a phone call.
The phone is visible clearly when the proposed version is zoomed onto the person, while it was too small for most users to notice in the scaled video.
This resulted in a 3.5 times higher $IQ$ value (Table \ref{table:results-quantativeC}, Question 1).
Similarly, in video D, most users who viewed the scaled video failed to see a cup in the person's hand.
This lead to a low $IQ$ value of 0.08 (Table \ref{table:results-quantativeD}, Question 1).
On the other hand, 47\% of the users who viewed the proposed version of D correctly stated that the person is holding a cup.
This resulted in a 5.9 times higher $IQ$ value.
Note that one major task of a security operator is to understand the activities at the surveillance site, and recognizing small objects carried by people in the scene is crucial for situation assessment.
From this, we conclude that sZoom produces videos that are more appropriate for surveillance.
Averaging over both videos and all questions, sZoom improves the information quotient by 99\%.

\begin{table}
\centering
\caption{Information Quotient ($IQ$) for the three questions about video C}
\begin{tabular}{|m{5.5cm}|c|c|}\hline
\multicolumn{1}{|c|}{\multirow{2}{*}{Question}} & Scaled  & \textbf{sZoomed} \\
& ($IQ$)& ($IQ$) \\
\hline
1. What is the person by the car doing in the beginning (first ten seconds) of the video? &0.22 &\textbf{0.77}\\
\hline
2. What is the person with the dolly doing? A dolly is a small platform on wheels used for carrying heavy objects. &0.33 & \textbf{0.65}\\
\hline
3. How many people does the largest group of people consist of? People standing close to each other form a group. & 0.5 & \textbf{0.88}\\
\hline
\multicolumn{1}{|c|}{\textbf{Average}} & 0.35 & \textbf{0.77}\\
\hline
\end{tabular}
\label{table:results-quantativeC}
\end{table}

\begin{table}
\centering
\caption{Information Quotient ($IQ$) for the six questions about video D.}
\begin{tabular}{|m{5.5cm}|c|c|}\hline
\multicolumn{1}{|c|}{\multirow{2}{*}{Question}} & Scaled  & \textbf{sZoomed} \\
& ($IQ$)& ($IQ$) \\
\hline
1. What is the right person of the group of two in the beginning of the video holding in his hand? &0.08 &\textbf{0.47}\\
\hline
2. How many people in the video are wearing a hat? &0.31 & \textbf{0.58}\\
\hline
3. What does the person in the red shirt waiting by the posts do when he sees his friend arriving by car? & 0.61 & \textbf{0.68}\\
\hline
4. What is the person in the black shirt doing while getting out of the car? & 0.30 & \textbf{0.68}\\
\hline
5. How many groups of people are walking through the video?  & 0.46 & \textbf{0.68}\\
\hline
6. Where are the two people (purple shirt and orange/purple striped shirt) before they walk through the scene? & 0.46 & \textbf{0.79}\\
\hline
\multicolumn{1}{|c|}{\textbf{Average}} & 0.37 & \textbf{0.65}\\
\hline
\end{tabular}
\label{table:results-quantativeD}
\end{table}

\subsection{Comparison with the Previous Version}
To the best of our knowledge, the work by Kuang et al.~\cite{kuang2014real} is the only previous state of the art work on automatic zoom in surveillance videos. As mentioned earlier, sZoom extends the previous work in terms of adaptive refinement of the zoom window, improved fusion framework, and improved penalty mechanism. With these improvements, sZoom creates videos that are more suitable for surveillance. In order to demonstrate the improvements, we repeated the experiment of Section \ref{Qualitative User Study} with two versions of the video: sZoomed and PV ( preliminary version, which is also the state of the art \cite{kuang2014real}). We used two 1 minute videos that were used in previous work \cite{kuang2014real} from the VIRAT dataset~\cite{oh2011large}. The web link to the user study can be found here\footnote{https://sites.google.com/site/szoomuserstudy/}. A total of 28 users (18 male, 10 female) participated in the study. They were aged between 18 and 32. The results of the experiment are given in Table \ref{table:new-study}. 

\begin{table*}
\centering
\caption{Comparison of the proposed method with the previous work (PV). The table includes average ratings for the five questions and standard deviation in parentheses.}
\begin{tabular}{|m{5cm}|m{1.3cm}|m{1.3cm}|m{1.3cm}|m{1.3cm}|m{1cm}|m{1cm}|}\hline
\multicolumn{1}{|c|}{\multirow{2}{*}{Statement}}&\multicolumn{2}{c|}{Video A}& \multicolumn{2}{c|}{Video B} &\multicolumn{2}{c|}{t-Test}
\\\cline{2-7}
& Scaled  & \textbf{sZoomed}  &Scaled  & \textbf{sZoomed}& $t_{stat}$ & $t_{cric}$ \\
\hline
1: The video is able to provide details of the regions that are important for surveillance. &2.7 (1)	&		\textbf{4.14} (0.84)	&		3.28 (0.84)	&		\textbf{4.12} (1.01)	& 7.3 & 1.68\\
\hline
2: The video provides a detailed view of all important events and activities happening at the site. &	2.89 (1.2)	&		\textbf{3.96} (1.07)	&		3.2 (1.0)	&		\textbf{4.08} (0.86) & 5.66 & 1.68	\\
\hline
3: The presentation of the video reduces boredom and keeps me engaged in the activities in the video. &	2.53 (1.4)	&		\textbf{4.1} (0.95)	&		3.56 (1.08)	&		\textbf{4.0} (0.91) & 5.34 & 1.68\\
\hline
4: The presentation of the video is helpful in monitoring the area. &2.6 (1.1)	&		\textbf{4.03} (0.83)	&		3.52 (0.91)	&		\textbf{4.2} (0.91)	& 7.7 & 1.68\\
\hline
5: It is easy to understand the activities in the video.	 &3.07 (1.2)	&		\textbf{4.2} (0.91)	&		3.48 (1.04)	&		\textbf{4.28} (0.91) & 7.4 & 1.68	\\
\hline
\multicolumn{1}{|c|}{\textbf{Average}} & 2.77 (1.24)	&		\textbf{4.1} (0.92)	&		3.4 (0.98)	&		\textbf{4.12} (0.89)	\\
\hline
\end{tabular}
\label{table:new-study}
\end{table*}

For both the videos, we notice significantly higher ratings for the video produced by sZoom as compared to the previous work. Between both PV videos, Video 1 received much lower ratings. This is because the face detector falsely detected a pole as a face in the previous work. In the proposed framework, this problem is handled by the fusion framework through a conformal coefficient. Overall, from Table \ref{table:results-qualitative} and Table \ref{table:new-study}, it is apparent that the scaled version achieves the lowest, and the proposed work the highest rating, with the previous work lying in-between. We chose to keep the tables separate because they are the results of two different user studies with different set of users. It can still be seen that the users consistently rated the sZoomed video the highest with a score around 4 out of 5. To analyze the statistical significance of the results, we conducted one-tailed t-test across both videos. 

The degree of freedon was 52 and the alpha value was 0.05. As we see in the last two columns of Table \ref{table:new-study}, the t statistic is significantly higher than t critical. The probability of observing the data with the null hypothesis of equal means was less than 0.00001 for all five questions. This proves that the results are not by chance; the proposed work has higher ratings than the previous work.

\subsection{Limitations}
The proposed framework is mainly applicable during day time. During night time the analysis algorithms may not be able to detect the threats. In order to use the framework in night videos, specialized threat detectors are required that work on night IR video. Another limitation of the framework is single device support. In the current framework, we need to create a new stream for each device dimension. However, it is not a severe limitations as there are only few standard display sizes of the mobile devices.

\section{Literature Review}\label{Related Work}
The presented work is most strongly connected to \emph{video retargeting}. Video retargeting typically deals with high resolution input videos that were created for presentation on large screens. The goal is to change the size or the aspect ratio of such a video to fit a smaller screen size, e.g., the display of a mobile device. As much of the important content of the video as possible should remain visible in the retargeted version. At the same time, the retargeting process should keep the amount of introduced artifacts like deformed objects to the minimum. Temporal stability is another desirable goal for the retargeted video.
The retargeting parameters should remain consistent from one frame to the next in order to not introduce temporally varying artifacts like shakiness or flicker.

More specifically, this paper presents a cropping-based retargeting approach. In this family of techniques, a rectangular area is chosen in every frame, such that it contains the highest amount of visual importance for a given size
~\cite{chen2016learning,gaddam2015cameraman,koccberber2014video,li2010video}. All pixels outside of the rectangle are discarded entirely. The choice of a suitable rectangle is a compromise between losing detail through scaling the video and losing context through cropping. The locations of cropping windows in each frame over time describe a trajectory through the space-time cube of the video.
This trajectory is constrained by cinematographic rules for panning that require a certain amount of smoothness, and an adequate amount of zoom needs to
be chosen~\cite{knoche2007kindest}. An optimal trajectory may be determined in a number of different ways. If multiple optimal trajectories are found, artificial cuts may be introduced to transition between them. The advantage of cropping-based video retargeting is that the content in the chosen area of interest is preserved faithfully, and few visual artifacts are introduced.
A common disadvantage, however, is that important content may be discarded entirely.

Importance-based scaling (also called ``warping'') is another popular group of retargeting approaches~\cite{wang2011scalable,gallea2014physical}. Instead of cutting out important content, the frames are divided into smaller
image areas which are then scaled down separately according to their respective
importance. The goal is to keep the size and shape of important foreground objects nearly identical, while less important background areas are shrunk and may be deformed. This distributes the loss of detail and deformation over those regions of a frame that the viewer is paying the least attention to.
Warping approaches are very powerful and are able to produce visually pleasing
results for still images. When applying them to videos, maintaining temporal stability during highly dynamic scenes is a major challenge.

Seam Carving is a technique that was originally developed for the retargeting of
still images~\cite{avidan2007seam}. It works by finding a vertical seam of connected pixels that crosses the image from top to bottom.
The seam is chosen so that its pixels cover the minimum possible amount of
visual importance. Removing the pixels belonging to the seam from every row of the image reduces the width of the image by one. The process of finding and removing seams can be iterated to reduce the width by arbitrary values. The height can be decreased analogously by removing horizontal seams. Due to its tendency to remove pixels from unimportant regions, Seam Carving has
been reported to produce excellent results for images with large unstructured
areas such as the sky, water or walls.
However, problems may occur when the shape of the unstructured area bears
importance to the image, or when an image mainly consists of structured
backgrounds and straight lines.
Seam Carving has also been adapted for the purpose of retargeting videos~\cite{yan2013matching}.
Here, the Seam Carving technique may be applied to each frame of a video
individually while adding temporal consistency as another optimization criterion.

Approaches based on both warping or Seam Carving are generally unsuited in a
video surveillance scenario.
This is mainly due to the modification of the content of the video.
Deforming objects in the scene or changing their position and size relative to
each other may alter the context and lead to misinformation.
Furthermore, the potential of introducing artificial motion through temporal
instability is undesirable in a surveillance scenario where object motion is an
important cue.
In general, all video retargeting approaches -- including the cropping-based
ones -- focus on the aesthetics of the results.
The output videos should be pleasant to watch~\cite{xiang2010video}.
Completely removing certain content may be desirable over partially cutting off
an object, and deformation in low attention background areas may be tolerable.
In our surveillance scenario, aesthetics are only a minor goal.
It may be sacrificed to achieve a more faithful preservation of content and
context.
The composition of the scene must not be tampered with.
It is also imperative that there are no ``blind spots'' in the video where
content has been removed permanently.
Everything that is contained in the original video should also appear in the
retargeted video at some point in time.

The basis for most retargeting approaches is a metric that allows to measure the amount of visual attention that is given to objects in the scene by a human observer.
If the content has previously been watched by many different people and their viewing behavior was recorded, the problem of measuring visual attention can be solved by crowd-sourcing~\cite{carlier2010crowdsourced}.
However in most cases, such information is unavailable.
By exploiting the fact that humans pay more attention to objects that are salient, i.e. have a high contrast to their surrounding, visual attention can instead be estimated by determining saliency.
Algorithms to determine saliency often mimic the way the human visual system perceives contrast.
The contrast can be calculated between a center region and its surrounding~\cite{liu2011learning,mahadevan2010spatiotemporal} or between segmented regions of the image~\cite{cheng2011global,ren2014region}.
Additionally, some approaches consider contrast on multiple different scales~\cite{liu2011learning,goferman2012context,guo2010novel}.
Motion is also an important feature to consider when calculating the saliency in videos.
If the motion of an object is in contrast with the motion of its surrounding, the object is perceived as salient~\cite{mahadevan2010spatiotemporal,guo2010novel}.

Note that no metric of visual attention can perfectly model the amount of sensitivity of an image area in a surveillance scenario as it would be perceived by a well-trained and attentive operator.
To compensate for this discrepancy, our system frequently zooms out to a full view of the entire scene.
Errors in the sensitivity estimation can thus never \emph{completely} discard important content.

From an application perspective, our work falls into the category of mobile video surveillance systems. 
Architecturally, such systems are composed of source modules, image processing and sink nodes.
Source modules sense the environment and send their data to an image processing module.
The data may now be analyzed using techniques like person re-identification~\cite{tao2013person} or crowded scene analysis~\cite{licrowded}.
A sink node can be an alarm or a display.
The term \emph{mobile} in such a system may refer to mobility of any of its components~\cite{cucchiara2010mobile}.
In our scenario, the source module is a stationary camera.
All signal processing is stationary as well and connected to network infrastructure.
The sink nodes are mobile phones connected through a mobile data network.
Mobile video surveillance from an architectural framework perspective with mobility in the same components as in our case is presented in~\cite{chung2012smartphone} and~\cite{zhang2012design}.
With their focus on components and communication protocols without addressing adaptation to limited screen size, the scope of these papers differs from ours.

\emph{Online object tracking} means to find an object's location and shape in the frames of a video where only the initial position and appearance of the object are known.
\emph{Online} refers to the circumstance that the location is updated frame by frame and there is no prior knowledge of the entire video.
Tracking algorithms can be characterized by the features that are used to describe and detect the object, the search mechanism (e.g., particle filters), and how the target model is updated to compensate for appearance variation.
Each type of tracker has its individual strengths and weaknesses.
If the situation-dependent performance of a tracker can be measured, the tracking quality may be increased by fusing the outputs of multiple trackers~\cite{biresaw2014tracker}.
In our work, we use the well-established mean shift tracking approach to adjust the position of the zoomed-in area if the object of interest is moving~\cite{comaniciu2000real}.
As this paper does not contribute to the field of tracking, we refer the reader to a survey article that covers many generic online tracking algorithms~\cite{wu2013online}.

Table \ref{table:related-work} gives a comparative overview of the related works.
We compare the related approaches to our work by using the following criteria:
\begin{itemize}
\item \textbf{Online:} In a surveillance scenario, the full video is unavailable beforehand. The approaches must be applicable to a live stream of data.
\item \textbf{Scene coverage:} Security operators need to be able to see the entire scene and all surrounding context for situation assessment. No part of the scene must be left out for too long.
\item \textbf{Sensitivity focused on surveillance:} The sensitivity criteria by which important regions in a video are identified should be based on the needs of the surveillance tasks. 
\item \textbf{Unmodified content:} In a surveillance scenario, modification of the video content is undesirable. 
\item \textbf{Automatic RoI:} In a scene with multiple regions of interest, manually zooming into them is tiring. The framework should be capable of automatically selecting a region to display.
\end{itemize}

\begin{table}
	\centering
\caption{Summary of previous works.}
\scriptsize
		\begin{tabular}{|p{1cm}|p{0.6cm}|p{0.7cm}|p{1.2cm}|p{1.1cm}|p{1cm}|}
			\hline
			Previous Work &  Online & Scene coverage & Surveillance sensitivity & Unmodified content & Automatic RoI\\\hline
			Li et al. \cite{li2010video}          & no & no & no & yes & yes \\\hline
             Xiang et al. \cite{xiang2010video}       & no & no & no & yes & yes \\\hline
			Carlier et al. \cite{carlier2011combining} & no & yes & no & yes & no \\\hline
			Wang et al. \cite{wang2011scalable}     & no & no & no & no & yes \\\hline
			Pang et al. \cite{pang2011classx}       & yes & yes & no & yes & no \\\hline
			Shafiei et al. \cite{shafiei2012jiku}      & yes & yes & no & yes & no \\\hline
			Yan et al. \cite{yan2013matching}      & yes & no & no & no & yes \\\hline
			Koccberber et al. \cite{koccberber2014video}  & no & no & no & yes & yes \\\hline
			Gallea et al. \cite{gallea2014physical}   & yes & no & no & no & yes \\\hline
			Lim et al. \cite{lim2014isurveillance} & yes & yes & yes & yes & no \\\hline
			Wang et al. \cite{wang2015wireless}     & yes & yes & no & yes & no \\\hline
			Zhai et al. \cite{zhai2015object}       & no & yes & yes & yes & no \\\hline
			Gaddam et al. \cite{gaddam2015cameraman}  & yes & no & no & yes & yes \\\hline
			Chen et al. \cite{chen2016learning}     & yes & no & no & yes & yes \\\hline
			\textbf{Our work}                    &\textbf{yes} & \textbf{yes} & \textbf{yes} & \textbf{yes} & \textbf{yes} \\\hline
		\end{tabular}
\normalsize 
\label{table:related-work}
\end{table}

\section{Conclusions}\label{Conclusions}
In this paper we have proposed sZoom, an automatic zoom framework for surveillance videos. sZoom identifies sensitive regions in the high resolution input video and produces a low resolution retargeted video for devices with a smaller display.
By digitally zooming into sensitive regions, sZoom provides the security operator with a more detailed view.
The sensitivity of the regions in the scene is determined by fusing semantic observations, user input, and a multi-variate Gaussian penalty.
The final zoom is achieved with a cubic spline interpolation of pan and zoom with iterative refinement according to target ROI tracking.
Based on the experiments with a complete near real-time implementation of sZoom, we draw the following conclusions:

\begin{itemize}
\item The proposed zoom method assists the operators in monitoring with mobile devices by allowing them to gain 99\% more information compared to watching a scaled video even with the baseline detectors.
\item The sZoom produced video is most appropriate for surveillance task in comparison to the scaled version of the video and the state of the art \cite{kuang2014real}.
\item By tracking the ROI while zooming in, the system accuracy is improved by 42\% compared to the preliminary version of the system.
\item To reduce the processing time, the resolution of the input frame for foreground detection and body detection were reduced to 60\% and 80\% of full HD, respectively, with a negligible ($\approx 0.05$) decrease in F1 value.  
\item An accumulation window of 4 frames ($\omega=4$) is provides best the detection accuracy for all three detectors. The reduced window size and resolution dramatically reduced the processing time to 5\% of the original time taken by the baseline system. 
\end{itemize}

In the future, we want to extend the region selection across multiple cameras to create an online retargeted mash-up of multiple surveillance videos. The challenge lies in seamlessly switching from one camera to another without compromising the security operator's ability to assess longer lasting activities.

\bibliographystyle{IEEEtran}
\bibliography{reference}

\begin{IEEEbiography}[{\includegraphics[width=1in,height=1.25in,clip,keepaspectratio]{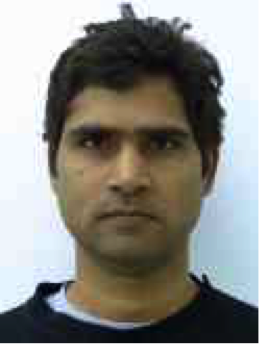}}]{Dr. Mukesh Saini} is an Assistant Professor at IIT Ropar. His research focus is broadly on innovative use and design of image processing and machine learning algorithms for multimedia systems. Areas of his recent work include surveillance, video authoring, and smart classrooms. In addition, he is also exploring multimedia techniques to solve safety related problems in the emerging area of Smart Cities and IoT.  Dr. Mukesh obtained Master of Technology (M. Tech) in Electronics Design and Technology from Indian Institute of Science (IISc), Bangalore, in 2006 and PhD in Computer Science from School of Computing, National University of Singapore, Singapore in 2012 respectively. He worked as post doctoral researcher at National University of Singapore, University of Ottawa, and New York University. He is an IEEE and ACM member.
Mukesh has played roles of Reviewer, TPC member, Tutorial organizer, and Panelist for various reputed conferences and journals. He is an editor of Multimedia Technical Committee (MMTC) Communication Reviews. 
\end{IEEEbiography}

\begin{IEEEbiography}[{\includegraphics[width=1in,height=1.25in,clip,keepaspectratio]{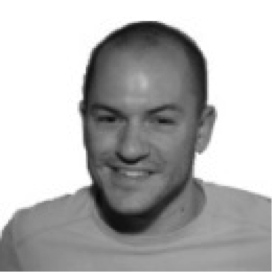}}]{Benjamin Guthier} received his PhD in computer science from the University of Mannheim, Germany in 2012, where his main research focus was the creation of high dynamic range video in real-time. Since 2012, he is working as a post-doc researcher at the University of Mannheim with a one year stay at the University of Ottawa, Canada as a guest researcher. His current research interests include image and video processing, smart cities and the extraction of information on affect from text.
\end{IEEEbiography}

\begin{IEEEbiography}[{\includegraphics[width=1in,height=1.25in,clip,keepaspectratio]{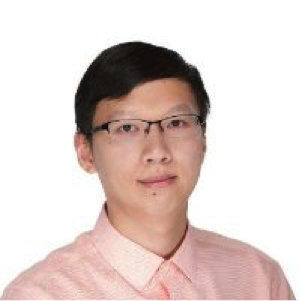}}]{Hao Kuang} has Master's from University of Ottawa. His research areas include multimedia, image processing , and augmented reality.
\end{IEEEbiography}

\begin{IEEEbiography}[{\includegraphics[width=1in,height=1.25in,clip,keepaspectratio]{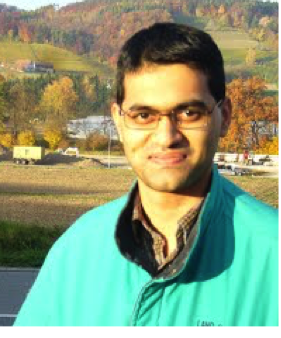}}]{Dwarikanath Mahapatra}  is currently a Research Staff Member at IBM Research Australia. He obtained his Ph.D. from the National University of Singapore in 2011, and worked as a post-doctoral research fellow at the ETH Zurich, Switzerland from 2011-2015. Dwarikanath's research interests are mainly in medical image analysis, machine learning, deep learning, decision support systems and computer aided diagnosis. He also explores other aspects of computer vision such as object detection and tracking, surveillance and image classification using deep neural networks. Dwarikanath has published more than 50 papers in prestigious conferences and journals. During his Ph.D. research he developed algorithms that use models of the human visual system for medical image analysis such as registration and segmentation. He also developed a novel method for detecting salient regions in videos which is used as a baseline for many course projects at the National University of Singapore. As part of his post-doctoral research at the ETH Zurich he developed algorithms and models for the diagnosis and detection of Crohn's disease using advanced machine learning techniques. In his current role at IBM he is involved in medical image analysis research to drive IBM Research's focus on healthcare research.
\end{IEEEbiography}

\begin{IEEEbiography}[{\includegraphics[width=1in,height=1.25in,clip,keepaspectratio]{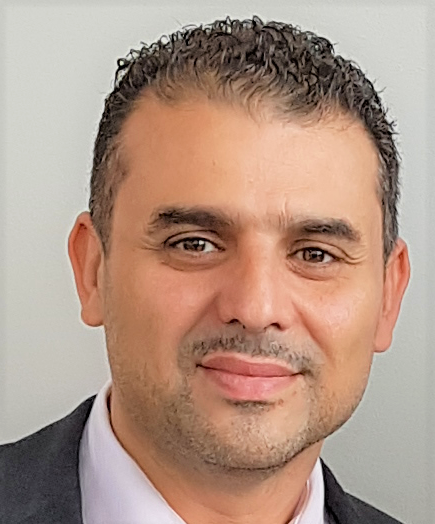}}]{Abdulmotaleb El Saddik} (M01 SM04 F09) is Distinguished University Professor and University Research Chair in the School of Electrical Engineering and Computer Science at the University of Ottawa. His research focus is on the establishment of Digital Twins using AI, AR/VR and Tactile Internet that allow people to interact in real-time with one another as well as with their digital representation. He has authored and co-authored 10 books and more than 550 publications and chaired more than 50 conferences and workshop. He has received research grants and contracts totaling more than \$18 M. He has supervised more than 120 researchers and received several international awards, among others, are ACM Distinguished Scientist, Fellow of the Engineering Institute of Canada, Fellow of the Canadian Academy of Engineers and Fellow of IEEE, IEEE I\&M Technical Achievement Award. IEEE Canada C.C. Gotlieb (Computer) Medal and A.G.L. McNaughton Gold Medal for important contributions to the field of computer engineering and science.
\end{IEEEbiography}

\EOD

\end{document}